\documentclass{vldb}
\usepackage{graphicx}
\usepackage{balance}
\usepackage{comment}
\usepackage{subcaption}
\usepackage{url}
\usepackage[hidelinks]{hyperref}
\vldbTitle{Understanding the Scalability of Hyperledger Fabric}
\vldbAuthors{Nguyen Minh Quang, Dumitrel Loghin, Tien Tuan Anh Dinh}
\vldbDOI{https://doi.org/10.14778/xxxxxxx.xxxxxxx}
\vldbVolume{12}
\vldbNumber{xxx}
\vldbYear{2019}

\pdfoutput=1

\begin{document}

% ****************** TITLE ****************************************
\title{Understanding the Scalability of Hyperledger Fabric}

% ****************** AUTHORS **************************************


\numberofauthors{3}

\author{
% 1st. author
\alignauthor
Minh Quang Nguyen\\
       \affaddr{National University of Singapore}\\       
       \email{quang.nguyen@u.nus.edu}
% 2nd. author
\alignauthor
Dumitrel Loghin \\
       \affaddr{National University of Singapore}\\       
       \email{dumitrel@comp.nus.edu.sg}
% 3rd. author
\alignauthor 
Tien Tuan Anh Dinh\\
       \affaddr{Singapore University of Technology and Design}\\
       \email{dinhtta@sutd.edu.sg}
}

\maketitle

\begin{abstract}

The rapid growth of blockchain systems leads to increasing interest in understanding and comparing blockchain
performance at scale.  
%growth in popularity of permissioned blockchain systems, performance
%benchmarking has become a primary concern toward analyzing the fundamental
%bottlenecks of various blockchain platforms and tackling unresolved scalability
%problem of their underlying distributed consensus protocols. 
In this paper, we focus on analyzing the performance of Hyperledger Fabric v1.1 --- one of the most
popular permissioned blockchain systems. Prior works have analyzed Hyperledger Fabric v0.6 in depth, but newer
versions of the system undergo significant changes that warrant new analysis. Existing works on benchmarking
the system are limited in their scope: some consider only small networks, others consider scalability of only
parts of the system instead of the whole.  

%As Fabric has moved its version to v1.1  bringing in
%elemental modifications to its architecture and consensus layer, which used to be the main scaling bottleneck
%for v0.6, there has been considerable research work benchmarking Fabric's performance for building real-world
%  blockchain applications. Nevertheless, none of them efficiently addresses the scalability of Hyperledger
%  Fabric v1.1 under practical settings or examines the scaling bottlenecks of Fabric, which are indeed the
%  major concerns for transitioning the technology into production
%phase.

We perform a comprehensive performance analysis of Hyperledger Fabric v1.1 at scale. We extend an existing
benchmarking tool to conduct experiments over many servers while scaling all important components of the
system. Our results demonstrate that Fabric v1.1's scalability bottlenecks lie in the communication overhead
between the execution and ordering phase. Furthermore, we show that scaling the Kafka cluster that is used for
the ordering phase does not affect the overall throughput.

\end{abstract}

\section{Introduction}
 
Blockchain technology is moving beyond cryptocurrency
applications~\cite{blockcoin,Bitcoin}, and becoming a new platform for general
transaction processing.  A blockchain is a distributed ledger which is
maintained by a network of nodes that do not trust each other. At each node, the
ledger is stored as a chain of blocks, where each block is cryptographically
linked to the previous block.  Compared to a traditional distributed database
system, a blockchain can tolerate stronger failures, namely Byzantine failures
in which malicious nodes can behave arbitrarily.

There are two types of blockchains: permissionless and permissioned. The performance of both types of systems,
however, lags far behind that of a typical database. The primary bottleneck is the consensus protocol used to
ensure consistency among nodes. Proof-of-Work~\cite{Bitcoin}, for example, is highly expensive and achieves
very low throughputs, whereas PBFT~\cite{pbft} does not scale to a large number of nodes~\cite{blockbench}.
Beside consensus, another source of inefficiency is the order-execute transaction model, in which transactions
are first ordered into blocks, then they are executed sequentially by every node. This model, adopted by
popular blockchains such as Ethereum and Hyperledger Fabric v0.6, is not efficient because there is no
concurrency in transaction execution~\cite{solo-benchmark}. 

New versions of Hyperledger Fabric, namely version v1.1 and later, implement a
new transaction model called execute-order-validate  model. Inspired by
optimistic concurrency control mechanisms in database systems, this model
consists of three phases. In the first phase, transactions are executed (or
simulated) speculatively. This simulation does not affect the global state of
the ledger. In the second phase, they are ordered and grouped into blocks.
In the third phase, called validation or commit, they are checked for conflicts
between the order and the execution results. Finally, non-conflicting
transactions are committed to the ledger. 

The ordering phase is performed by an ordering service which is loosely-coupled
with the blockchain. Hyperledger Fabric offers two types of ordering service:
{\em Solo} which is used for development and testing, and {\em Kafka} service
which is used for deployment in production system. The Kafka service forwards
transactions to an Apache Kafka cluster for ordering~\cite{kafka}. By allowing
parallel transaction execution, Hyperledger Fabric v1.1 can potentially achieve
higher transaction throughputs than systems that execute transactions
sequentially. However, it introduces an extra communication phase compared to
the order-execute model, thus incurring more overhead.

%The advantages of this model lie in the parallel transaction execution and,
%hence, the scaling capability. On the other hand, there is a concerns that
%Hyperledger Fabric v1.1 incurs overhead of extra steps in the transaction flow
%and communication between network components due to separating the ordering
%service from the peers. In this paper, we address this concern by analyzing the
%performance of Hyperledger Fabric v1.1 with Kafka ordering on a realistic
%cluster environment.

In this paper, we aim to provide a comprehensive performance analysis of the
execute-order-validate transaction model. To this end, we evaluate the
throughput and latency of Hyperledger Fabric v1.1 in a local cluster of up to 48
nodes, running with Kafka ordering service. Our work differs from
Blockbench~\cite{blockbench}, which benchmarks an earlier version of Hyperledger
Fabric with order-execute transaction model. Its scope is more extensive than
other recent works that examine the performance of Hyperledger Fabric v1.1 and
later. For example, \cite{hpfabric} does not consider the effect of scaling the
Kafka cluster on the performance. Similarly, \cite{gauge, solo-benchmark,
fabric_benchmark} fix the size of the Kafka cluster, and use fewer than 10
nodes. In contrast, we examine the impact of scaling Kafka cluster on the
overall performance, using up to 48 nodes in a local cluster.

Hyperledger Caliper~\cite{caliper} is the official benchmarking tool for
Hyperledger Fabric. However, it offers little support and documentation on how
to benchmark a real distributed Fabric setup with Kafka ordering service. Most
of the documentation and scripts are considering the Solo orderer and a single
client. To overcome this, we developed Caliper++ by enhacing Caliper with a set
of scripts to configure, start and benchmark a distributed Fabric network with
variable number and type of nodes.

%Fabric v1.1
%only considered endorsing peers in their experiments, except Parth Thakkar et al.\cite{fabric_benchmark} only
%benchmarked a fixed small Fabric network topology without scaling peers.  A. Baliga et al.\cite{gauge} did
%address scaling endorsing peers in Fabric network, but they only benchmarked Fabric systems under Solo
%ordering service and used a simple smart contract not simulating real-world transactional workloads.

%This paper presents a comprehensive study on the scalability performance of
%various Kafka-based Fabric network topologies and focuses on the bottlenecks of
%Fabric system. 
In summary, our main contributions are as follows:
%\begin{comment}
%The smart contract used in all of our experiments is Smallbank
%\cite{smallbank}, a highly suitable smart contract for benchmarking blockchain
%at scale as it is representative of the large class of transactional workloads
%such as TPC-C.
%\end{comment} 

\begin{itemize}
	
\item We extend the Caliper, Hyperledger's benchmarking tool, by adding support
for distributed benchmarking. The result is Caliper++, a benchmarking tool that
can start Fabric with varying sizes and configurations.

\item We perform a comprehensive experimental evaluation of Fabric v1.1 using
Smallbank smart contract. We scale the number of peers involved in all three
transaction phases of the execute-order-validate model, by using up to $48$
nodes in a local cluster.

\item We show that endorsing peer --- the one that perform the execute and
validate phase --- is the primary scalability bottleneck. In particular,
increasing the number of endorsing peers not only incurs overheads in the
execute and order phase, but also leads to degraded performance of the Kafka
ordering service. On the other hand, we observe that scaling the Kafka cluster
does not impact the overall throughput.

\end{itemize}

Section~\ref{sec:background} discusses the background on blockchain systems, and
the architecture of Hyperledger Fabric v1.1. Section~\ref{sec:rel_work}
describes related works on blockchain benchmarking.
Section~\ref{sec:caliper++} describes our benchmarking tool called Caliper++.
Section~\ref{sec:analysis} presents the experiment setup and results, before
concluding in Section~\ref{sec:conclusion}.

\section{Background}
\label{sec:background}

Blockchain networks can be classified as either permisionless (or public) or permissioned (or private). In the
former, such as Ethereum \cite{ethereum} and Bitcoin \cite{Bitcoin}, any node can join the network, can issue
and execute transactions.  In the latter, such as Hyperledger Fabric, Tendermint \cite{tendermint}, Chain
\cite{chain}, or Quorum \cite{quorum}, nodes must be authenticated and authorized to send or execute
transactions. 

The ledger stores all historical and current states of the blockchain in the
form of blocks linked together cryptographically. To append a block, all nodes
must agree. More specifically, the nodes reach agreement by running a
distributed consensus protocol. This protocol establishes a global order of the
transactions. The majority of permissionless blockchains use computation-based
consensus protocols, such as Proof-of-Work (PoW), to select a node that decides
which block is appended to the ledger. Permissioned blockchains, on the other
hand,  use communication-based protocols, such as PBFT~\cite{pbft}, in which
nodes have equal votes and go through multiple rounds of communication to reach
consensus. Permissioned blockchains have been shown to outperform permissionless
ones, although its performance is much lower to that of a database
system~\cite{blockbench}.
Nevertheless, they are useful for multi-organization applications where
authentication is required but participating organization do not trust each
other.

%The permissioned blockchain networks, which align with our interests in this
%work, provide a pragmatic solution to enterprise applications where: (a)
%authentication is necessary for participation of multiple organizations; (b)
%access control limits the issue-transaction right to only authorized nodes in
%the network, and provides administrative management; (c) consensus on a
%transaction is made with the participation of all nodes (i.e. all
%organizations), in which nodes have equal votes; (d) secure transaction is
%required.

\subsection{Hyperledger Fabric v1.1}
Hyperledger Fabric v1.1 (or Fabric) is a permissioned (or private) blockchain
designed specifically for enterprise applications. A blockchain smart contract,
also called chaincode in Fabric, can be implemented in any programming language.
A Fabric network comprises four types of nodes: endorsing nodes (or peers),
non-endorsing nodes (or peers), clients, and ordering service nodes. These nodes
may belong to different organizations. Each node is given an identity by a
Membership Service Provider (MSP), which is run by one of the organizations.

\textbf{Endorsing and Non-Endorsing Peers.} A peer in the system stores a copy
of the ledger in either GolevelDB \cite{goleveldb} or CouchDB \cite{couchdb}
database. It can be an endorsing peer if specified so in the \textit{Endorsement
Policies}; otherwise it is non-endorsing. Endorsing peers maintain the
chaincode, execute transactions, and create (or endorse) transactions to be
forwarded to the ordering nodes.
% the chaincode logic and executes it to endorse a transaction in the execute
% step.

\textbf{Endorsement Policies.} A endorsement policy is associated with a
chaincode, and it specifies the set of endorsing peers. Only designated
administrators can modify endorsement policies.

\textbf{System Chaincodes.} In addition to running chaincode specified by users,
Fabric peers run a number of pre-define system chaincodes. Theare are four
system chaincodes: the life cycle system chaincode (LSCC) for installing,
instantiating and updating chaincodes, the endorsement system chaincode (ESCC)
for endorsing transactions by digitally signing them, the validation system
chaincode (VSCC) for validating endorsement signatures, and the configuration
system chaincode (CSCC) for managing channel configurations.

\textbf{Channel.} Fabric supports multiple blockchains that use the same
ordering service. Each blockchain is identified by a channel, where members may
consist of different sets of peers. Transactions on a channel can only be viewed
by its members. The order of transactions in one channel is isolated from those
of in another channel and there is no coordination between channels
\cite{hpfabric}.

\textbf{Ordering Service.} The Ordering Service consists of multiple Ordering
Service Nodes (OSNs). The OSNs establish a global order of transactions and
construct blocks to be broadcast to peers. A block is created when one of the
following conditions is satisfied: (1) the number of transactions reaches a
specified threshold, (2) a specified timeout is reached; (3) the block size
reaches a specified threshold. Endorsing peers receive blocks directly from the
ordering service, while non-endorsing peers get blocks via a gossip protocol
from other endorsing peers and from the ordering service. A Solo ordering
service consists of a single node (or orderer) which serves all clients. A Kafka
ordering service consists of a Kafka cluster to which OSNs forward transactions
for ordering.

\textbf{Client.} A client sends transactions to endorsing peers, wait until it
receives all endorsement from these peers, then broadcasts the endorsed
transactions to OSNs.

\textbf{Transaction.} Separating the ordering service from the peers,
Hyperledger Fabric v1.1 adopts the execute-order-validate (also called
simulate-order-commit) model for executing a transaction. In contrast,
Hyperledger Fabric v0.6 and other blockchain systems use the order-execute
transaction model. In the next section, we describe the life cycle of a
transaction in Fabric v1.1.

\begin{figure}[tbp]
\centering
\includegraphics[width=0.45\textwidth]{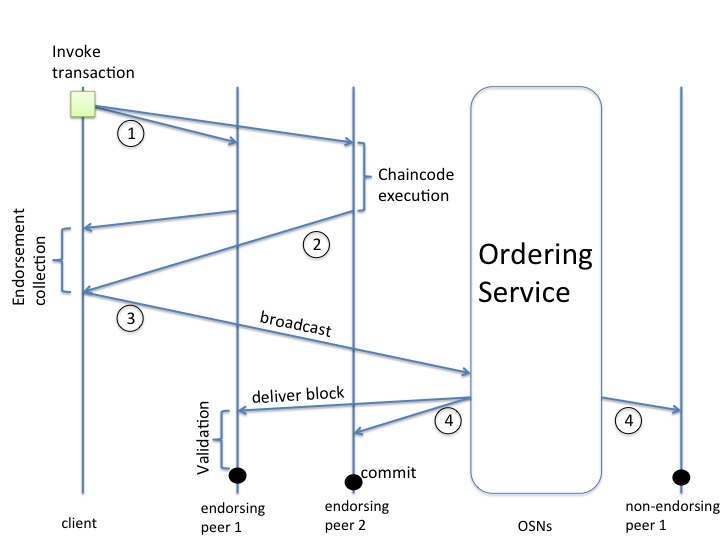}    
\caption{Transaction life cycle in Fabric v1.1.}
\label{transaction_flow}
\end{figure}

\subsection{Transaction Life Cycle in Fabric} 
Fabric v1.1 employs the novel execute-order-validate transaction model which
comprises three phases, as depicted in Figure~\ref{transaction_flow}.

\textbf{Endorsement phase.} A client submits a signed transaction to the
endorsing peers. Each endorsing peer verifies if the client is authorized to
invoke the transaction, then speculatively executes the transaction against its
local blockchain state. This process is done in parallel without coordination
among endorsing peers. Output of the execution, which consists of a read set and
a write set, is used to create an endorsement message. The peer signs the
endorsement and sends it back to the client.

\textbf{Ordering phase:} After collecting enough endorsements according to the
endorsement policy, the client creates an endorsed transaction and forwards it OSNs. The ordering service
orders the transactions globally. It then creates blocks and send them directly to endorsing peers or
gossiping them to non-endorsing peers.  

\textbf{Validation Phase:}  When receiving a block, every peer validates transactions in the block against the
endorsement policy of the chaincode. After
that, for every transaction, the peer checks for read-write conflict by comparing the key versions in the read
set are the same as those in the current ledger states. Any transaction that fails the validation or conflict
check is marked as invalid. Invalid transactions are discarded and their effects to the blockchain states are
rolled back. Finally, the block is appended to the peer's local ledger.

\section{Related Work}
\label{sec:rel_work}
There has been considerable interest in benchmarking Hyperledger Fabric. Blockbench \cite{blockbench} is the 
first framework for benchmarking permissioned blockchains. It divides blockchain stack into four layers:
consensus, data, execution and application. It contains many micro and macro benchmarks for evaluating
performance of every layer. However, Blockbench only supports Fabric v0.6 whose architecture is highly
different to that of v1.1. Although the results reported in~\cite{blockbench} cannot be extended to the new
system, they present useful baseline performance of the order-execute transaction model.  

E. Androulaki et al.\cite{hpfabric} evaluate Fabric v1.1 using Kafka ordering service. They use a Kafa-based
network of 5 endorsing peers, 4 Kafka nodes, 3 orderers, 3 Zookeeper nodes, and a varying number of
non-endorsing peers.  We note that non-endorsing peers do not play an active role in the transaction life
cycle, as shown in Figure~\ref{transaction_flow}. Therefore, they are not a potential scalability bottleneck.
We argue that the number of non-endorsing peers is not an important system parameter. As a consequence,
\cite{hpfabric} falls short in the analysis of system scalability.   

Parth Thakkar et al.\cite{fabric_benchmark} also benchmark Fabric v1.1. However, their network is small,
consisting of only 8 peers, 1 orderer and a Kafka cluster. The Kafka cluster runs on a single node, which does
not fully capture the communication overhead in a real system. Furthermore, they do not consider scalability
in terms of the number nodes. Ankur Sharma et al.\cite{solo-benchmark} use the Solo orderer which is not meant
to run in a real system. The network in~\cite{solo-benchmark} is small, with 4 peers, 1 client and 1 orderer
distributed on 6 cluster nodes. Similarly, A. Baliga et al.~\cite{gauge} use the Solo orderer. Although they
examine the effect of scaling the number of endorsing peers, they use a simple smart contract instead of ones
representing realistic transactional workloads like Smallbank or YCSB.   

%E. Androulaki et al.\cite{hpfabric} fixed a Kafka-based Fabric network of  5 endorsing 
%peers, 4 Kafka brokers, 3 Fabric orderers, and 3 ZooKeeper nodes and performed 
%various experiments with the FABCOIN chaincode. Their work focused more 
%on the impact of peer CPU on the performance of Fabric. They also tested the effect of
%scaling non-endorsing peers and reported that the system could scale up to a large number 
%of non-endorsing peers. However, non-endorsing peers do not participate in most of the 
%transaction flow , thereby not representing a potential scaling bottleneck of Fabric. 

\section{Caliper++}
\label{sec:caliper++}

In this section, we present Caliper++\footnote{The source code and
documentation are available on GitHub at
\url{https://github.com/quangtdn/caliper-plus}}, our extension to Hyperledger
Caliper~\cite{caliper}, the official benchmarking tool for Hyperledger Fabric.

\subsection{Starting Up Fabric Network}
The main challenge of benchmarking Kafka-based Fabric network at scale lies in
starting up the distributed Fabric network. While Fabric v0.6 has only one type
of nodes, represented by the peers themselves, Fabric v1.1 has six types of
nodes: endorsing peer, non-endorsing peer, orderer, Kafka node (or broker),
Zookeeper node, and client. This means there are more system parameters to
consider. It also makes it more complex to automatically start up the network
with all these components. In fact, we found limited official documentation on
configuring a Kafka-based system. There are even fewer documentations on running
Fabric on multiple physical nodes.

%In particular, while Hyperledger Fabric v1.1 documentation provides
%network samples using Solo orderer, it provides limited instructions and no sample on deploying Kafka-based
%Fabric network due to the complication in integrating Kafka ordering service. Another challenge is to set up
%the Fabric v1.1 network on multiple physical nodes without comprehensive documentations.  Lastly, allocating
%resources for a medium to large scale Kafka-based Fabric network under fully distributed setting can be costly
%in terms of the number of nodes required. For  example, the most basic Kafka ordering service already requires
%8 nodes (3 Zookeeper nodes, 4 Kafka brokers, 1 orderer nodes) as the least requirement for guaranteeing
%fault-tolerance of the Kafka cluster.

At its current state, Caliper \cite{caliper} -- the official benchmarking tool
for Hyperledger blockchains -- only supports testing on a local environment with
a limited set of predefined Fabric network topologies, all of which use the Solo
ordering service. Caliper++ supports additional functionalities for benchmarking
Kafka-based Fabric network at scale. In particular, it provides a set of scripts
to (a) auto-generate configuration files for any Fabric network topology, (b)
bring up large-scale Kafka-based Fabric network across cluster nodes, (c) launch
additional benchmarking tools at the operating system level, such as
\textit{dstat, strace, perf}. In addition, it allows benchmarking with
distributed clients.

\subsection{Benchmark Driver} 
Caliper++ implements the role of the client in Figure~\ref{transaction_flow}. It can be configured to
issue a given number of transactions with a fixed rate. After sending one transaction, it schedules sending
the next transaction after such interval that ensures the specified transaction rate is met. Responses from
endorsing peers received during the execution phase are processed asynchronously in the client's main event
loop.   

%waits for responses
%from the endorsing peers, then sleeps for an amount of time (depending on the transaction rate) before sending
%the next transaction.  %After finishing sending all the transactions, it processes messages from endorsing
%peers, and collects statistics regarding throughput and latency.  

We note that Caliper++ implements a benchmark client (or driver) that is tightly coupled to the Fabric
transaction workflow. This is different to Blockbench driver, which is separated from the blockchain network.
In particular, Blockbench driver sends transactions via JSON APIs, waits for the transaction IDs from one of
the blockchain nodes, then sleeps an appropriate amount of time before sending the next transaction. This
driver is simply a workload generator, which is independent of the blockchain transaction processing workflow.
Caliper++, in contrast, implements specific logic in which it waits and validates for responses from endorsing
peers against endorsement policies. Another difference of Caliper++ is that it processes responses from
multiple nodes for each transaction, as opposed to from a single node as in Blockbench driver. 

%One advantage of the Caliper client's design is that it supports fixed firing
%request rate, which has been used for benchmarking Hyperledger Fabric v1.1 in
%recent work \cite{fabric_benchmark, gauge}, while the client used in Blockbench
%waits for the ID of transaction to return before sending the next transaction,
%which  results in potentially variable transaction rate.

\section{Performance Analysis}
\label{sec:analysis}

In this section, we present our extensive benchmarking-based analysis of
Hyperledger Fabric v1.1 using Caliper++. In this analysis, we focus on the
following important performance metrics: throughput, latency and scalability.
The throughput represents the number of successful transactions per second
(tps). This is the most common metric to evaluate the performance of distributed
systems, including distributed databases and blockchains. Latency represents the
response time per transaction, in seconds. In our experiments, latency is taken
as the average of the response time of all transactions. Scalability represents
the changes in throughput and latency as the Fabric network scales up.

\subsection{Experiments Setup}
\label{experiment_setup}

\begin{figure}[tbp]
\centering
\includegraphics[width=0.45\textwidth]{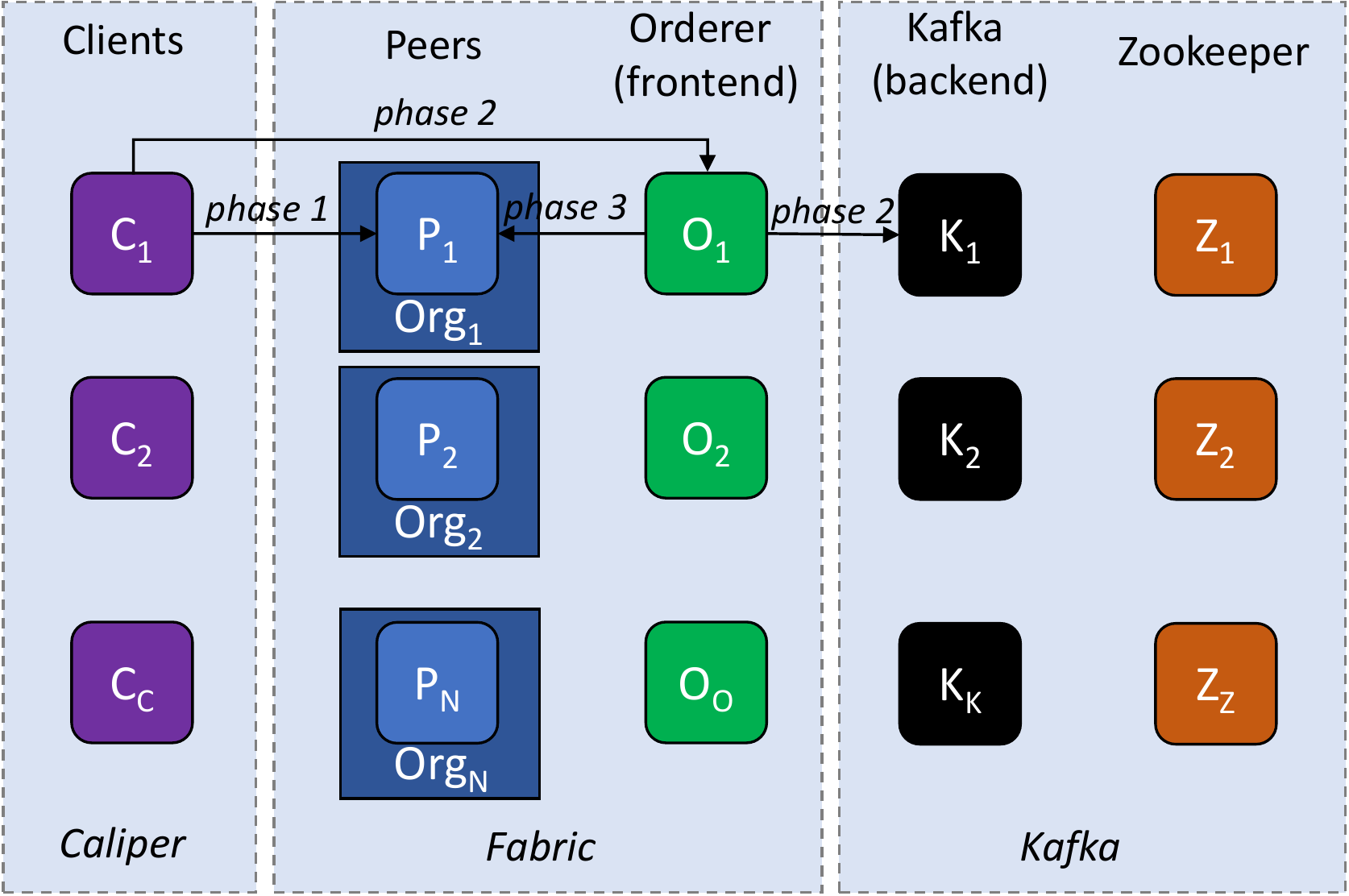}    
\caption{Hyperledger Fabric Setup}
\label{fabric_setup}
\end{figure}

In this paper, a Hyperledger Fabric network topology is described by the number
of endorsing peers (or peers, for simplicity) $N$, number of clients $C$ with
mutual transaction send rate $T$, number of Fabric orderers $O$, number of Kafka
brokers $K$ and number of Zookeeper nodes $Z$, as depicted in
Figure~\ref{fabric_setup}. All the peers belong to different organizations
(orgs), are endorsing peers and use GolevelDB as the state database.
The endorsement policy contains all $N$ endorsing peers. The block size is set
to 100 transactions/block, and the timeout is set to 2s.

\begin{figure}[t!]
\centering
\begin{subfigure}{0.45\textwidth}
\includegraphics[width=\textwidth]{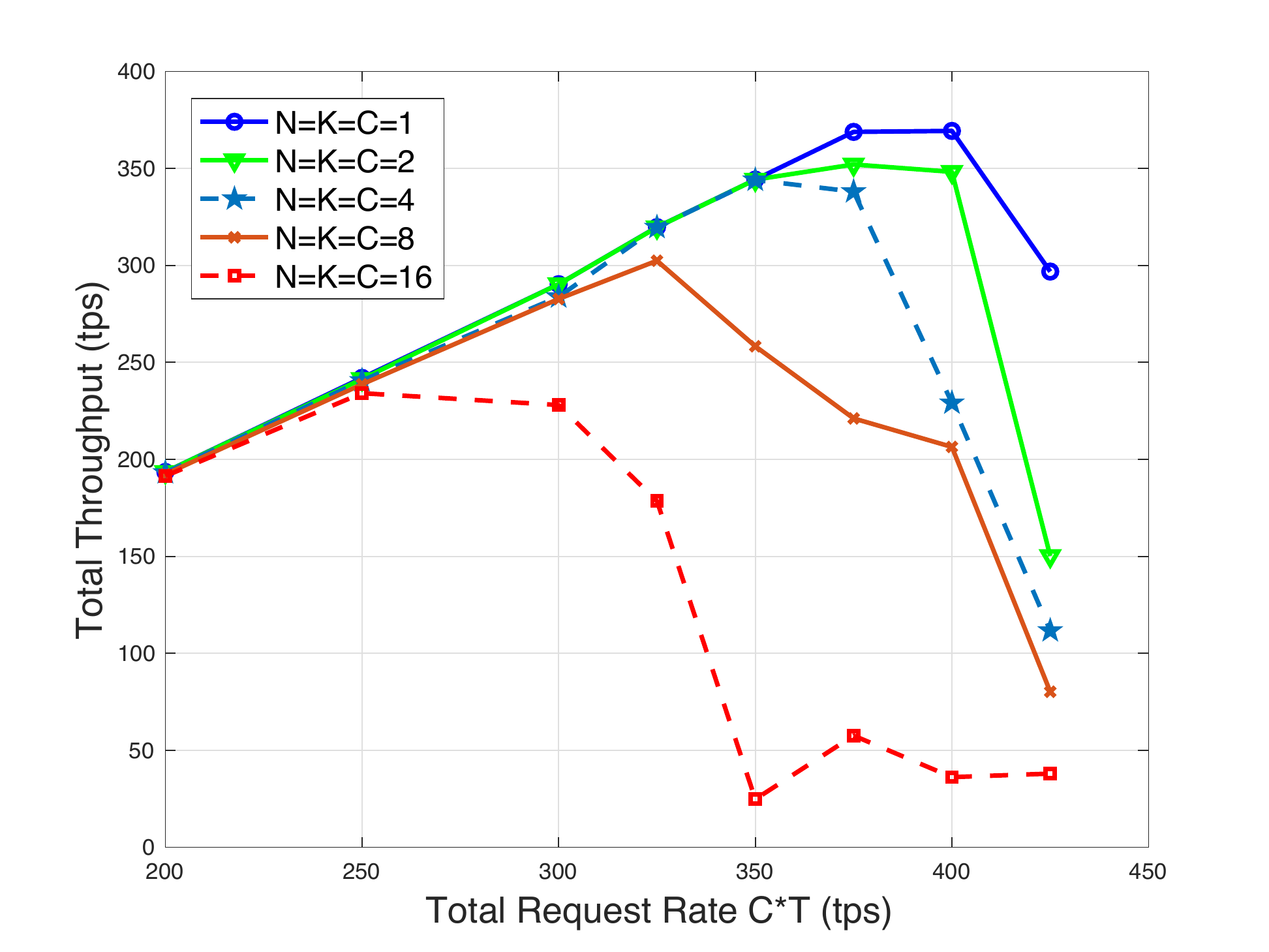}
\caption{Throughput for the setting $N=C=K$}
\label{saturation_point_throughput}
\end{subfigure}
\begin{subfigure}{0.45\textwidth}
\includegraphics[width=\textwidth]{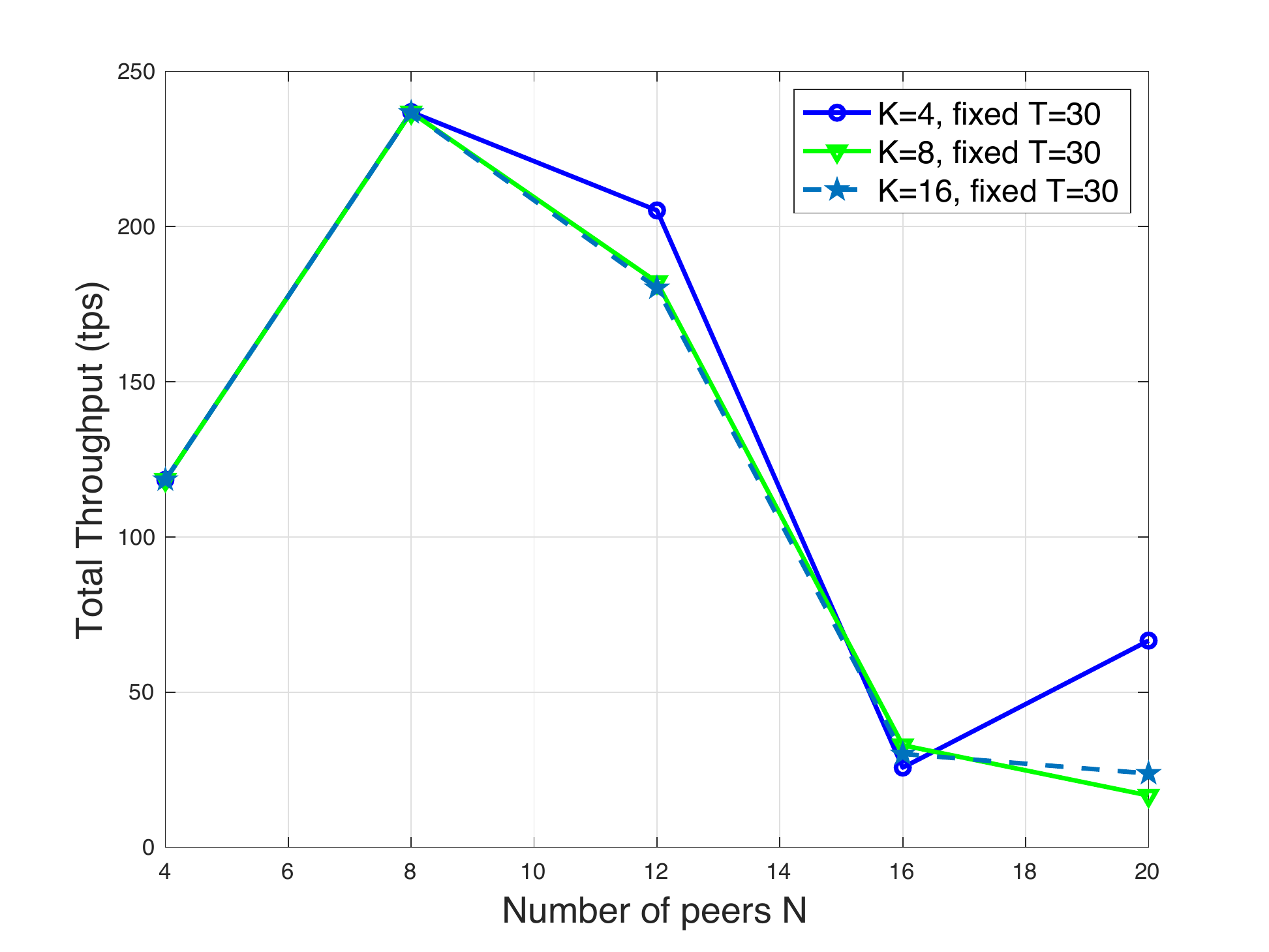}
\caption{Throughput for scaling clients with fixed rate $T=30$}
\label{fix_client_rate_throughput}
\end{subfigure}
\caption{Throughput while scaling network size}
\label{saturation_throughput}
\end{figure}

In a Kafka ordering service, the Fabric orderers only act as the proxies
forwarding transactions to the Kafka brokers which do the actual ordering.
Since only the number of Kafka brokers affects directly the ordering capacity of
the ordering service, we fix the number of Fabric orderers to $O=4$ throughout
our experiments and vary the number of Kafka brokers $K$ to assess the impact on
scaling Fabric. We shall see that increasing the number of Fabric orderers
strongly degrades the system throughput due to communication overhead.

For each Kafka broker, we set $min.insync.replicas = 2$, and
$default.replication.factor = K-1$. This configuration means that any
transaction is written on $K-1$ Kafka brokers, and committed after being
successfully written on two Kafka brokers. If $default.replication.factor <
K-1$, there would be idle Kafka brokers not participating in the ordering phase.
In practice, these idle Kafka brokers are reserved for ordering other channels
in the Fabric network. However, in our testing environment with single-channel
ordering service is not meaningful to have idle Kafka brokers. Furthermore,
setting a high value, such as $K-1$, for $default.replication.factor$ also
increases the fault-tolerance of the Kafka cluster, thereby providing us
insights on the trade-offs between performance and fault-tolerance.

We use the popular OLTP database benchmark workload Smallbank \cite{smallbank}
in our experiments. Simulating typical asset transfer scenario and a large class
of transactional workloads such as TPC-C in general, Smallbank is suited to test
the Fabric system at scale. Furthermore, Smallbank is one of the two
macro-benchmark workloads used in Blockbench
\cite{blockbench} to benchmark Hyperledger Fabric v0.6, and
thus gives us a stand for comparison between v0.6 and v1.1.

The experiments were run on a 48-node commodity cluster. Each server node has an
Intel Xeon E5-1650 CPU clocked at 3.5 GHz, 32 GB of RAM, 2 TB hard drive and a
Gigabit Ethernet card. The nodes are running Ubuntu 16.04 Xenial Xerus.

\begin{figure}[t!]
\centering
\begin{subfigure}{0.45\textwidth}
\includegraphics[width=\textwidth]{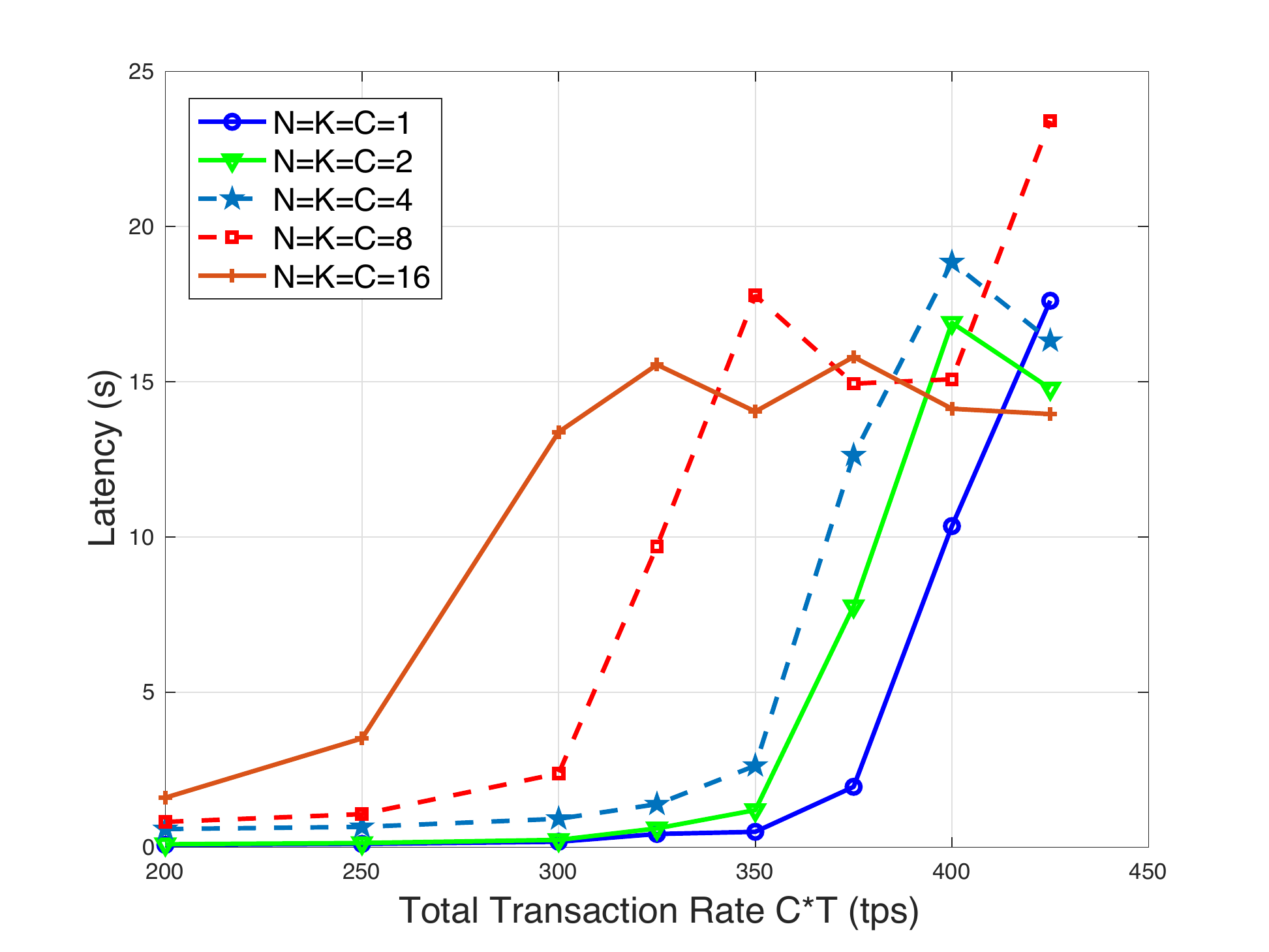}
\caption{Latency for the setting $N=C=K$}
\label{saturation_point_latency}
\end{subfigure}
\begin{subfigure}{0.45\textwidth}
\includegraphics[width=\textwidth]{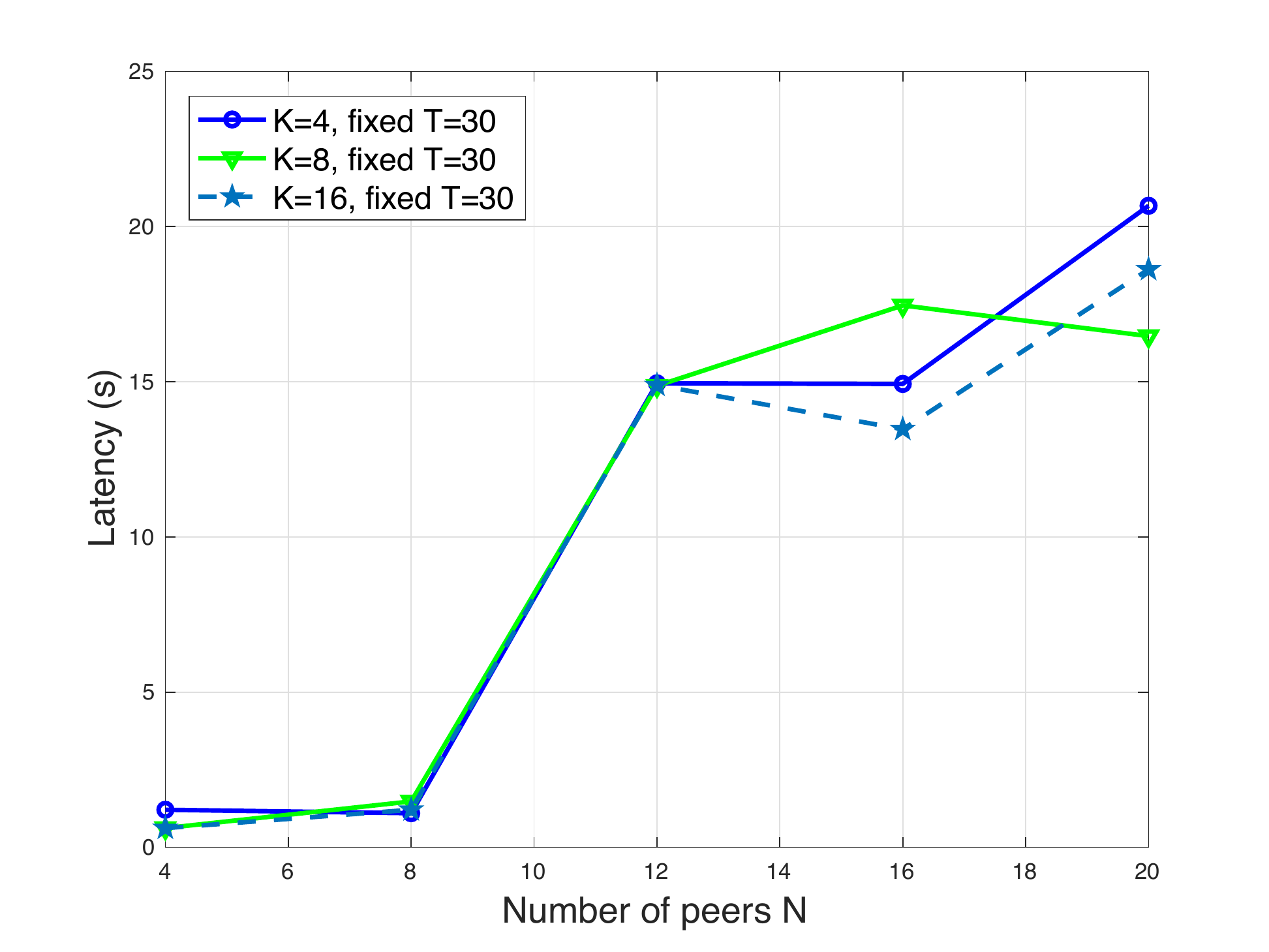}
\caption{Latency for scaling clients with fixed rate $T=30$}
\label{fix_client_rate_latency}
\end{subfigure}
\caption{Latency while scaling network size}
\label{saturation_latency}
\end{figure}

\subsection{Determining the Saturation Point}
\label{sec:saturation}

In this section, we examine the capacity of the Fabric system. We vary the
number of peers $N$, while keeping the number of clients equal to the number of
peers, $C=N$, and gradually increase the total request rate until the end-to-end
throughput is saturated. The saturation request rate $C*T$ determined in these
experiments is used in later experiments where we investigate the system's
performance while scaling the number of peers or Kafka brokers.

Firstly, we use a number of Kafka ordering brokers equal to the number of peers,
$K=N$, and incrementally increase the total request rate until the throughput
significantly degrades. The total throughput and latency are shown in Figure
\ref{saturation_point_throughput} and Figure \ref{saturation_point_latency},
respectively. We observe that the throughput degrades as the network scales up.
All network sizes exhibit throughput degradation on request rates equal or
higher than 400 tps. The network with 16 peers and Kafka orders is saturated
with a request rate of 300 tps. The latency exhibits similar behavior, except on
the 16 peers network where it starts to degrade from a request rate of 250 tps.
The fixed request rates of $C*T=300$ tps and $C*T=400$ tps represent, in
general, the points before and after the saturation, respectively. These request
rates are chosen to perform evaluate the costs of increasing the number of peers
or Kafka brokers.

\begin{figure}[t]
\centering
\includegraphics[width=0.47\textwidth]{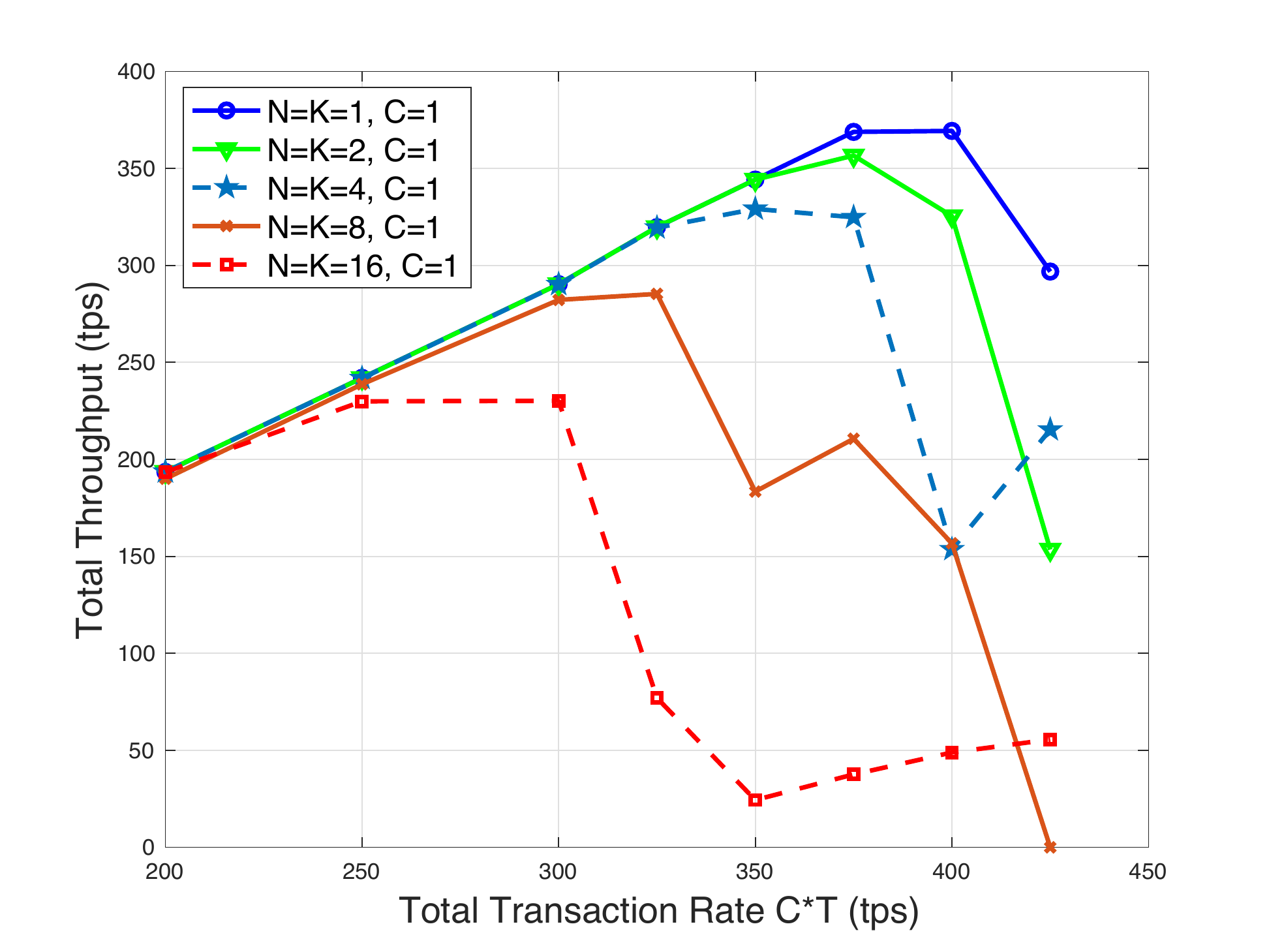}
\caption{Throughput with a single client $C=1$}
\label{saturation_point_throughput_1client}
\end{figure}

Secondly, we fix the number of Kafka brokers $K$ and the client's request rate
to $T=30$ , while scaling the number of peers and clients $N=C$. The network is
linearly enlarging in terms of both number of peers and total request rate.
The total throughput and latency are shown in Figure
\ref{fix_client_rate_throughput} and Figure \ref{fix_client_rate_latency},
respectively. We observe that the throughput decreases and the latency increases
when the network consists of more than 8 peers. Moreover, from $N=12$ on
all the tested network topologies, the throughput decreases drastically. The
cumulative request rate of the system at this point is 360 tps.

Thirdly, we show that the client does not represent a bottleneck by running the
experiments with a single client, $C=1$, while scaling the number of peers and
Kafka orderes, $N=K$. The throughput, depicted in
Figure~\ref{saturation_point_throughput_1client}, exhibits similar performance
as the throughput with more than one client.

We observe similar patterns in all the above experiments, exposing the same
saturation points in all Fabric network topologies under evaluation. This
suggests that the saturation of the networks is caused by Fabric's incapability
of handling request rates beyond certain thresholds, rather than the
communication or capacity bound of clients. Before the saturation point, the
throughput of the Fabric network with multiple clients is always slightly better
than that of Fabric network with a single client because, for the later, the
single client has to send a larger set of request transactions and accordingly
incurs more overhead.

\begin{figure*}[t] \centering
\begin{subfigure}{0.46\textwidth}
\includegraphics[width=\textwidth]{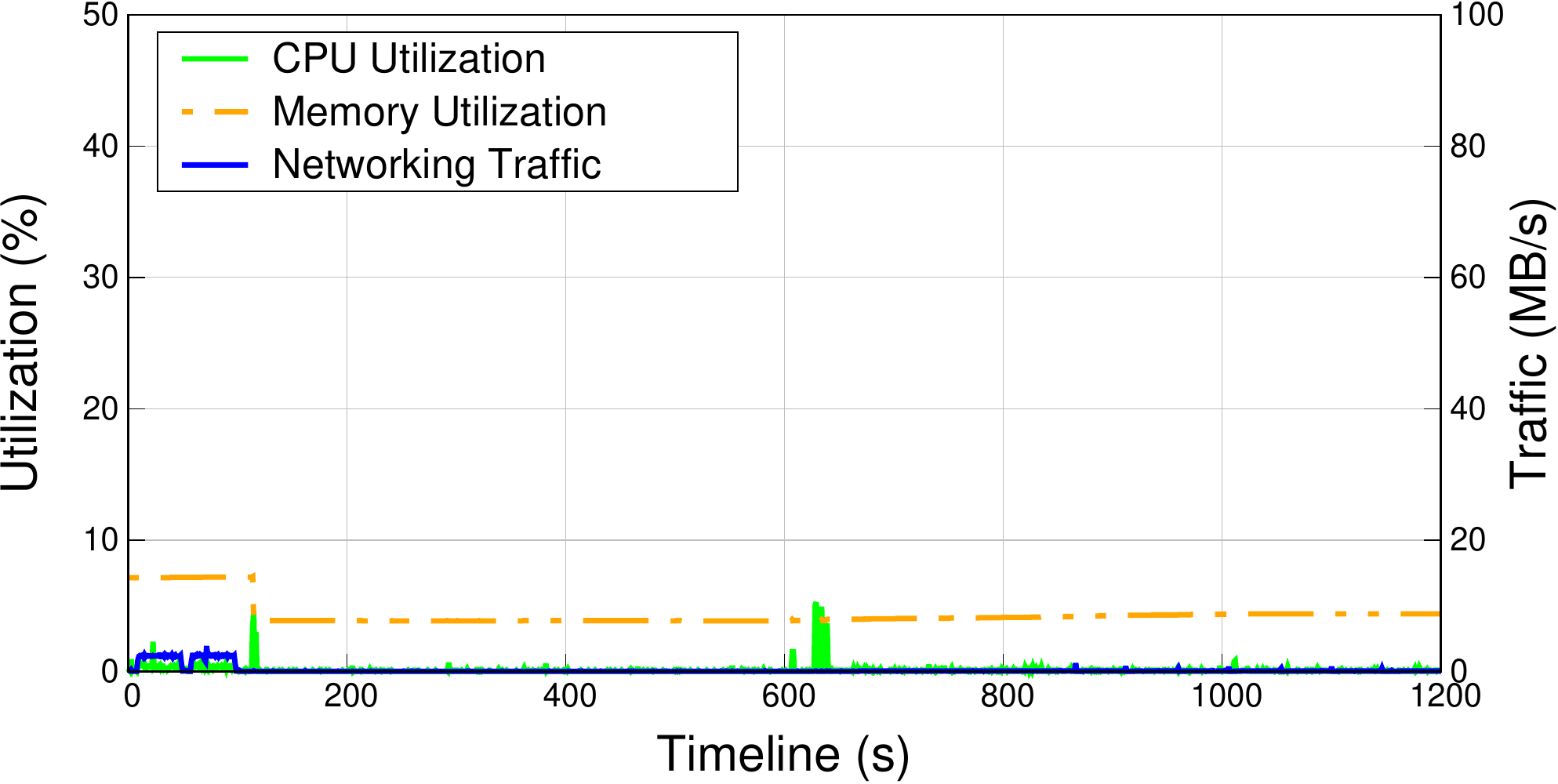}
\caption{Zookeeper}
\label{dstat_orderer}
\end{subfigure}
\begin{subfigure}{0.46\textwidth}
\includegraphics[width=\textwidth]{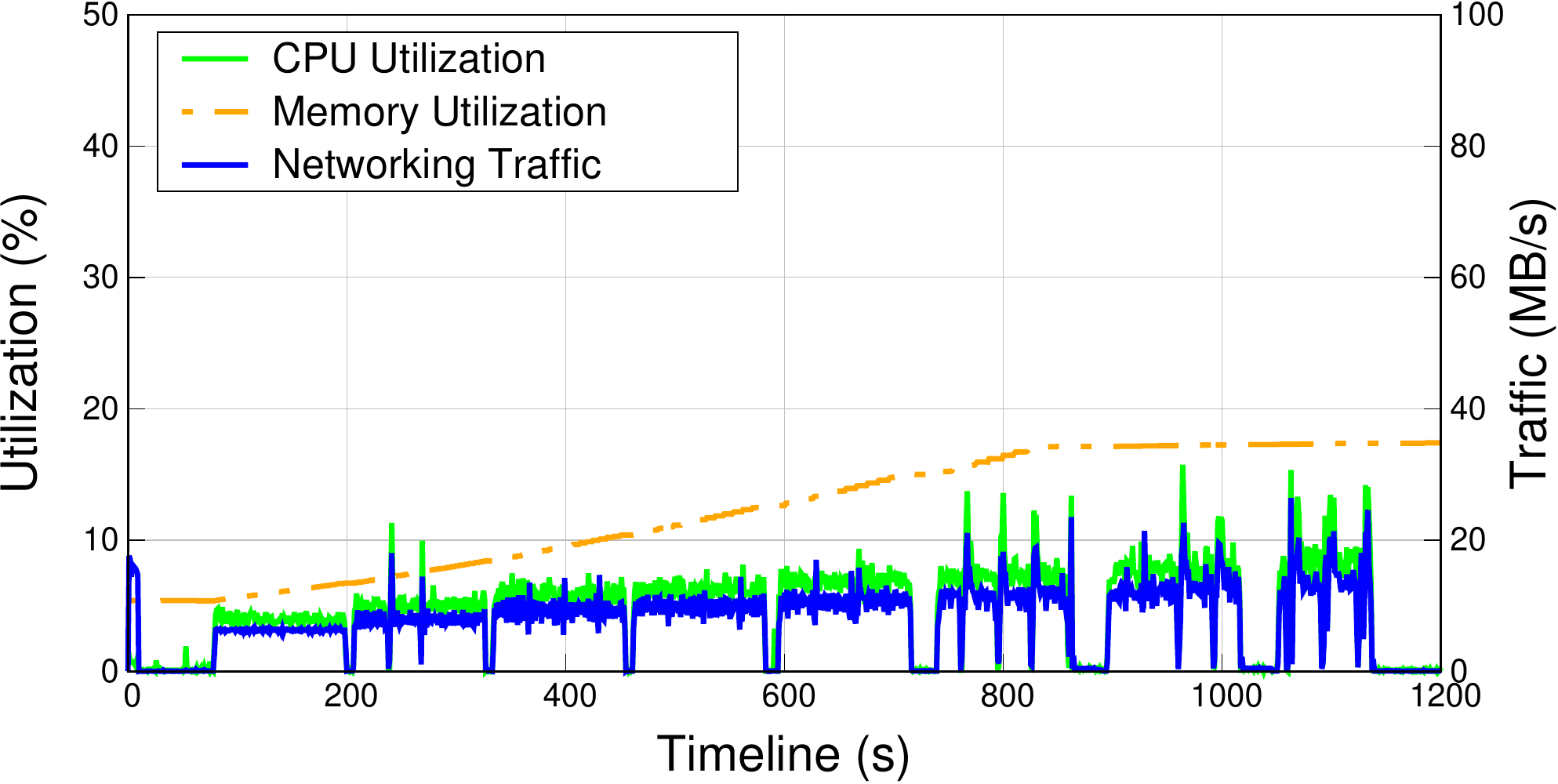}
\caption{Fabric Orderer}
\label{dstat_orderer}
\end{subfigure}
\begin{subfigure}{0.46\textwidth}
\includegraphics[width=\textwidth]{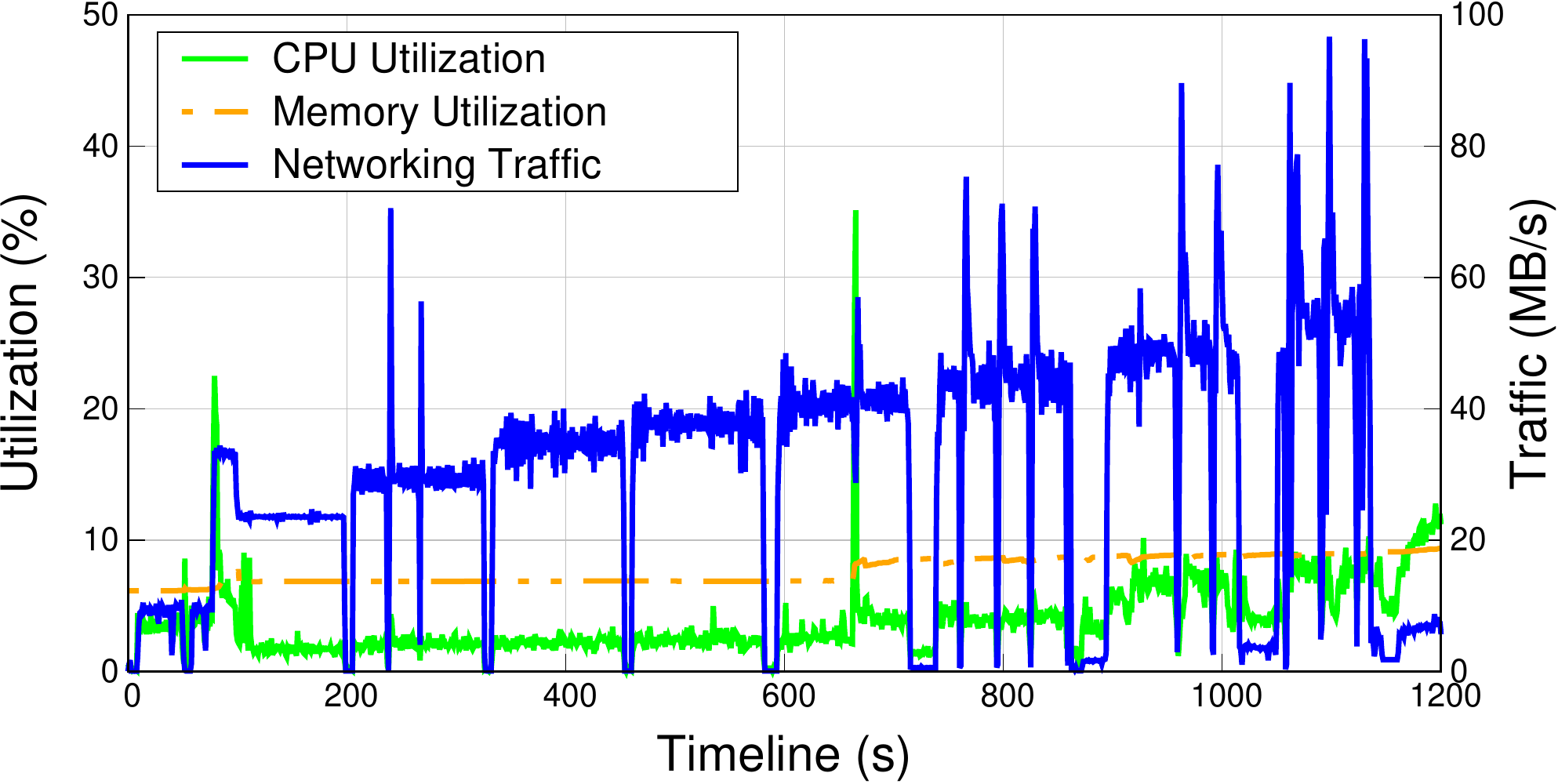}
\caption{Kafka Leader}
\label{dstat_kafka}
\end{subfigure}
\begin{subfigure}{0.46\textwidth}
\includegraphics[width=\textwidth]{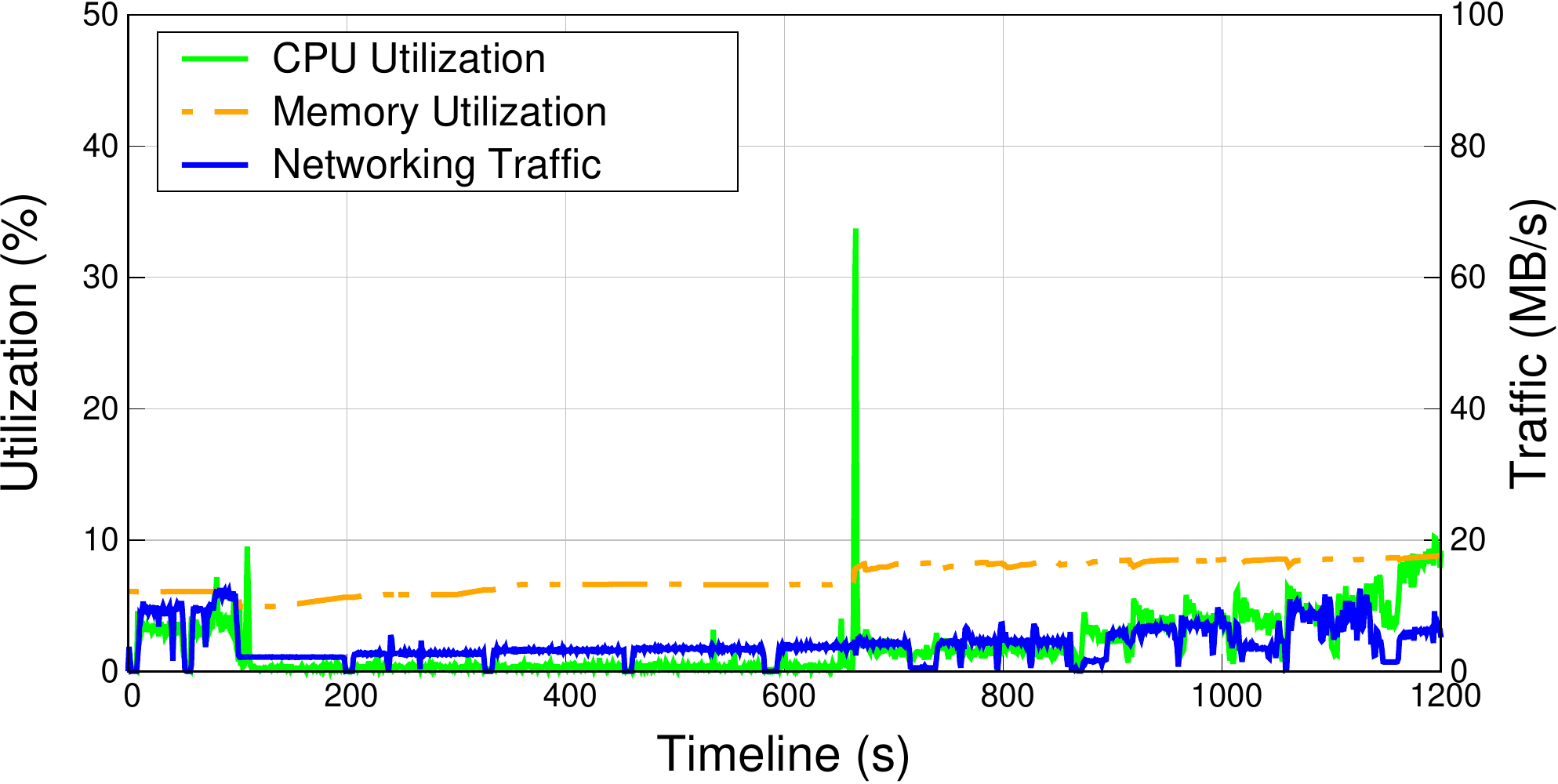}
\caption{Kafka Broker}
\label{dstat_kafka}
\end{subfigure}
\begin{subfigure}{0.46\textwidth}
\includegraphics[width=\textwidth]{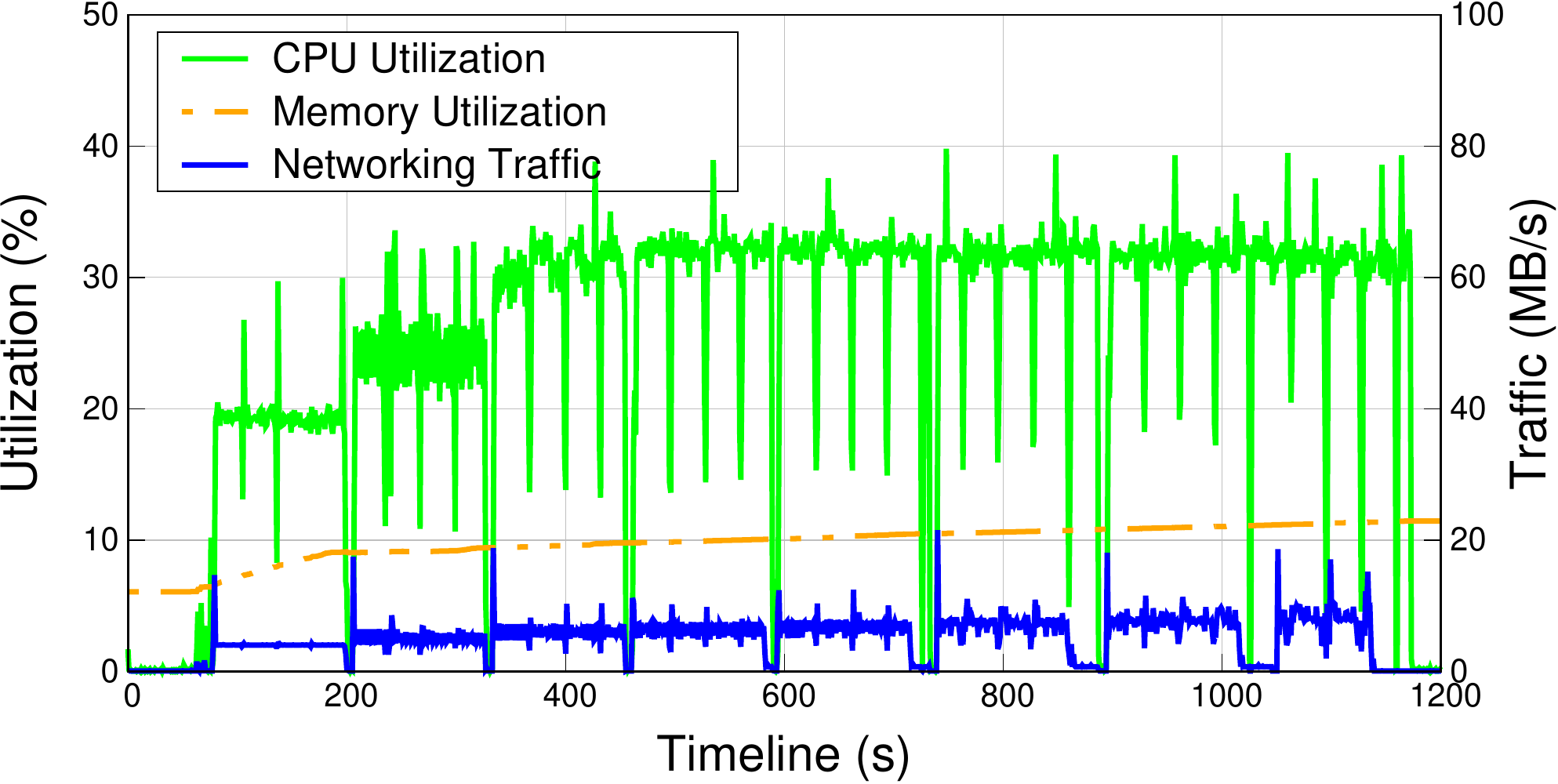}
\caption{Peer}
\label{dstat_peer}
\end{subfigure}
\begin{subfigure}{0.46\textwidth}
\includegraphics[width=\textwidth]{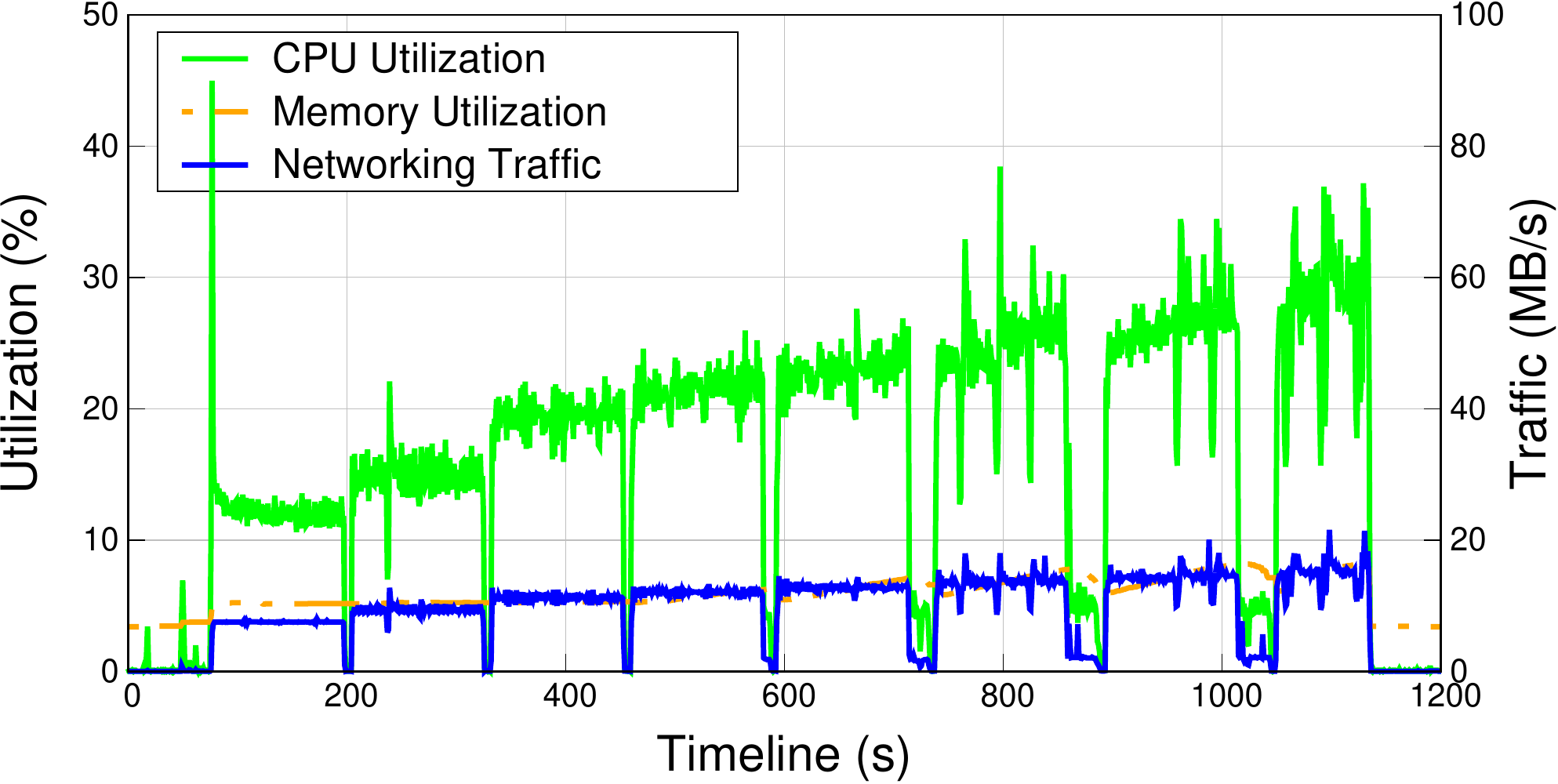}
\caption{Client}
\label{dstat_client}
\end{subfigure}
\caption{Resource utilization of different nodes in a Hyperledger Fabric
network}
\label{dstat}
\end{figure*}

To further validate our observation, we profile the systems with \textit{dstat}
tool to get CPU and memory utilization, and networking traffic on each node.
Figure~\ref{dstat} depicts the resource utilization of different types on nodes
in a Hyperledger Fabric network consisting of $N=8$ peers, $K=8$ Kafka nodes,
$O=4$ Fabric orderers, $Z=3$ Zookeeper nodes and $C=1$ client. The trends are
consistent with the benchmarking setting, where the clients use eight different
transaction request rates in the range $C*T \in [200,425]$. Each transaction
request benchmark lasts for 120s. We observe that client and peer resource
utilization, especially at CPU and networking level, increases with the
transaction rate, as expected.

However, none of the Fabric network nodes is fully utilizing the hardware. The
CPU utilization is below 40\% for all nodes. While the clients and the peers use
up to 30\% of the CPU, some Kafka brokers and Zookeeper nodes use less than 5\%.
Memory utilization is also low. Only the front-end orderer uses close to 20\%
(or around 6 GB) to buffer the requests.

\begin{figure}[t]
\centering
\includegraphics[width=0.47\textwidth]{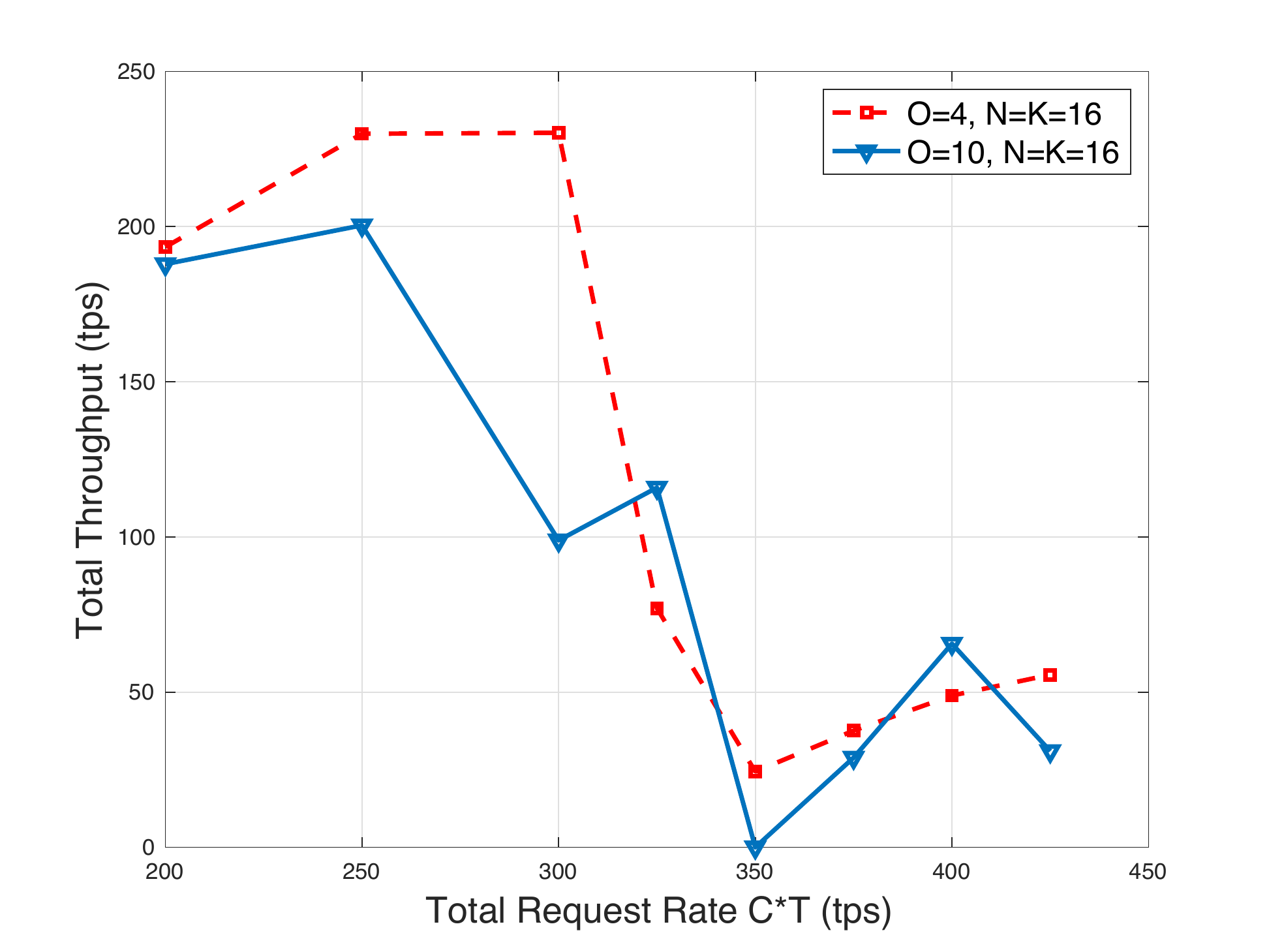}
\caption{Impact of increasing the number of orderers}
\label{varying_orderers}
\end{figure}

At networking level, the client, peers, Fabric orders and Kafka leaders exhibit
the highest traffic. This traffic represents both sent and received bytes. The
high traffic of Fabric orderer is explained by its double role, as receiver of
requests from the client and as dispatcher of requests to Kafka, as shown in
Figure~\ref{fabric_setup}. The high traffic of some Kafka nodes, or leaders, is
explained by their double role, as leaders managing the replication inside the
Kafka cluster and as connecting links to the Fabric orderers. Compared to the
maximum capacity of 125 MB/s of the Gigabit Ethernet, the traffic of Kafka
leaders is below 50\% or 60 MB/s, on average.

\subsection{Overhead of Fabric Orderers}

In this section, we investigate the impact of scaling the number of orderers. We
take the network size with the lowest performance in the previous section,
$N=K=16$, and increase the number of orderers from $O=4$ to $O=10$. The
throughput plotted in Figure~\ref{varying_orderers} shows that the network with
more orderers is performing worse. Increasing the number of orderers only
increases the communication overhead, since orderers act as proxies broadcasting
transactions rather than ordering.

\subsection{Scaling the Number of Fabric Peers}

In this set of experiments, we fix the request rate and investigate the impact
of increasing the number of peers. We fix the total request rate to $C*T=300$
and $C*T=400$, separately, as discussed in Section~\ref{sec:saturation}. We fix
the number of Kafka orderers on a value in the set $\{4, 8, 16\}$, and increase
the number of peers up to 24. The average throughput and latency of each
experimental setting is shown in Figure \ref{scaling_peer_throughput} and
\ref{scaling_peer_latency}, respectively.

\begin{figure}[t]
    \centering
    \begin{subfigure}{0.47\textwidth}
    \includegraphics[width=\textwidth]{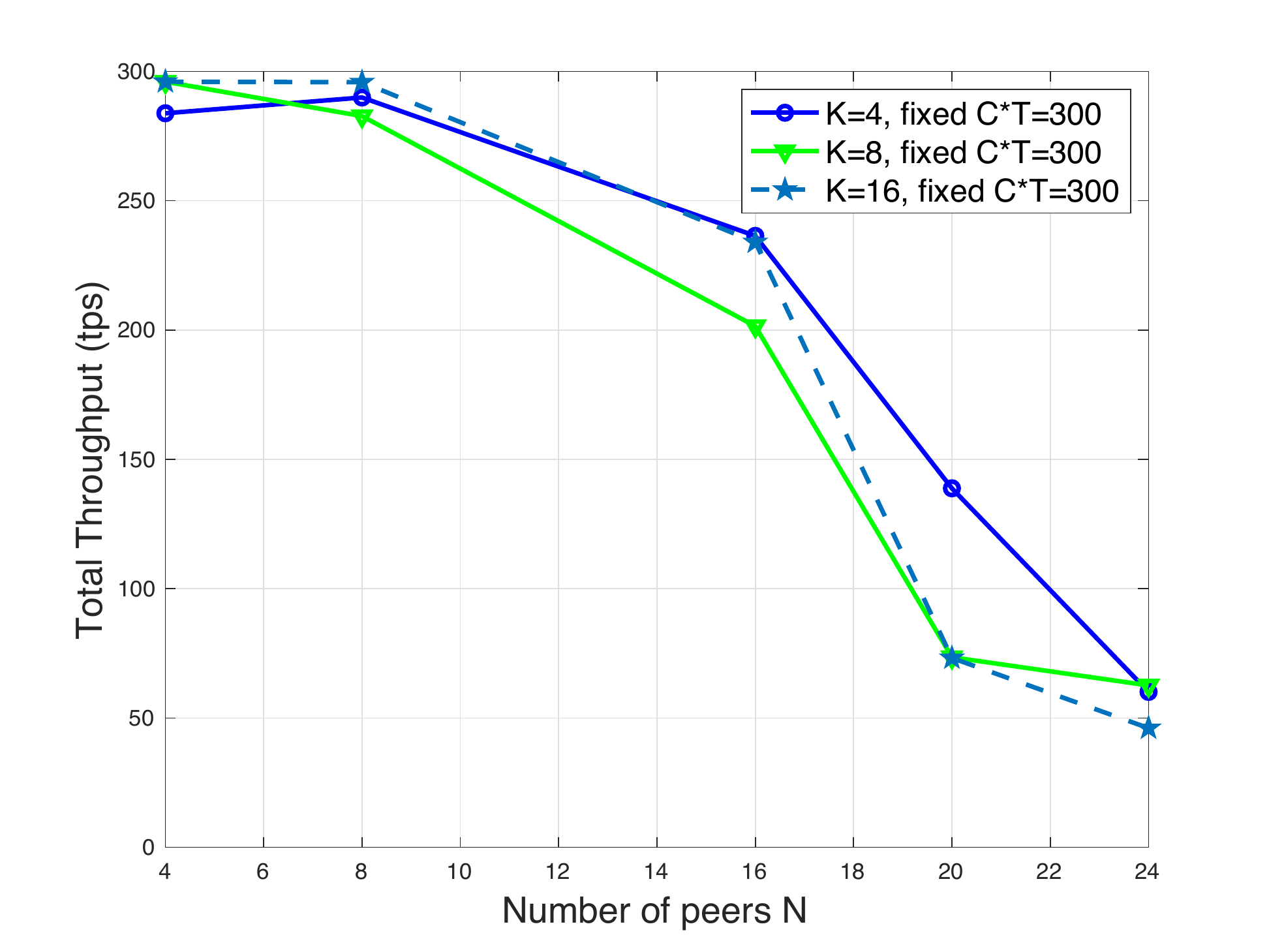}
    \caption{Throughput for fixed request rate $C*T=300$ tps}
    \label{scaling_peer_throughput}
    \end{subfigure}
    \begin{subfigure}{0.47\textwidth}
    \includegraphics[width=\textwidth]{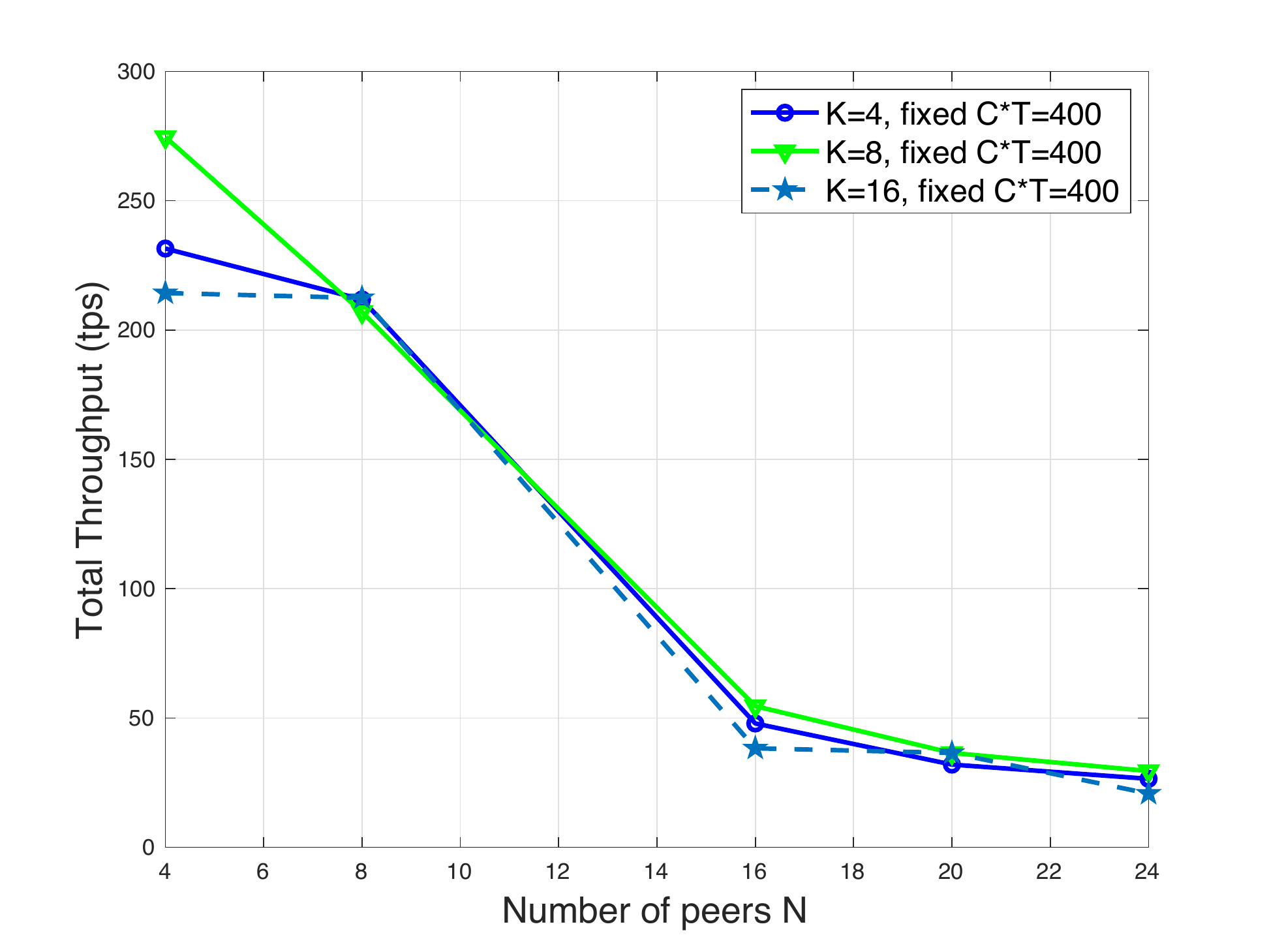}
    \caption{Throughput for fixed request rate $C*T=400$ tps}
    \label{scaling_peer_highthroughput}
    \end{subfigure}
    \caption{Throughput while scaling the number of peers}
    \label{scaling_peer_throughput}
\end{figure}

Increasing the number of peers strongly degrades the system's throughput and
limits Fabric' scalability. To understand the bottlenecks, we examine the system
logs and observe that  scaling the number of peers incurs overhead in both the
endorsement and ordering phases within the transaction flow.

Firstly, increasing the number of endorsing peers means that each client has to
wait for a larger set of endorsements from all the peers to prepare the endorser
transaction. Examining the logs of Caliper's clients, we observe that the
clients return a large number of timeout errors while collecting the
endorsements from peers and, accordingly, discard those transaction proposals.
As a result, clients send endorser transactions to the orderers at a much lower
rate due to dropped transactions and the time overhead for collecting
endorsements.

\begin{figure}[t]
    \centering
    \begin{subfigure}{0.49\textwidth}
    \includegraphics[width=\textwidth]{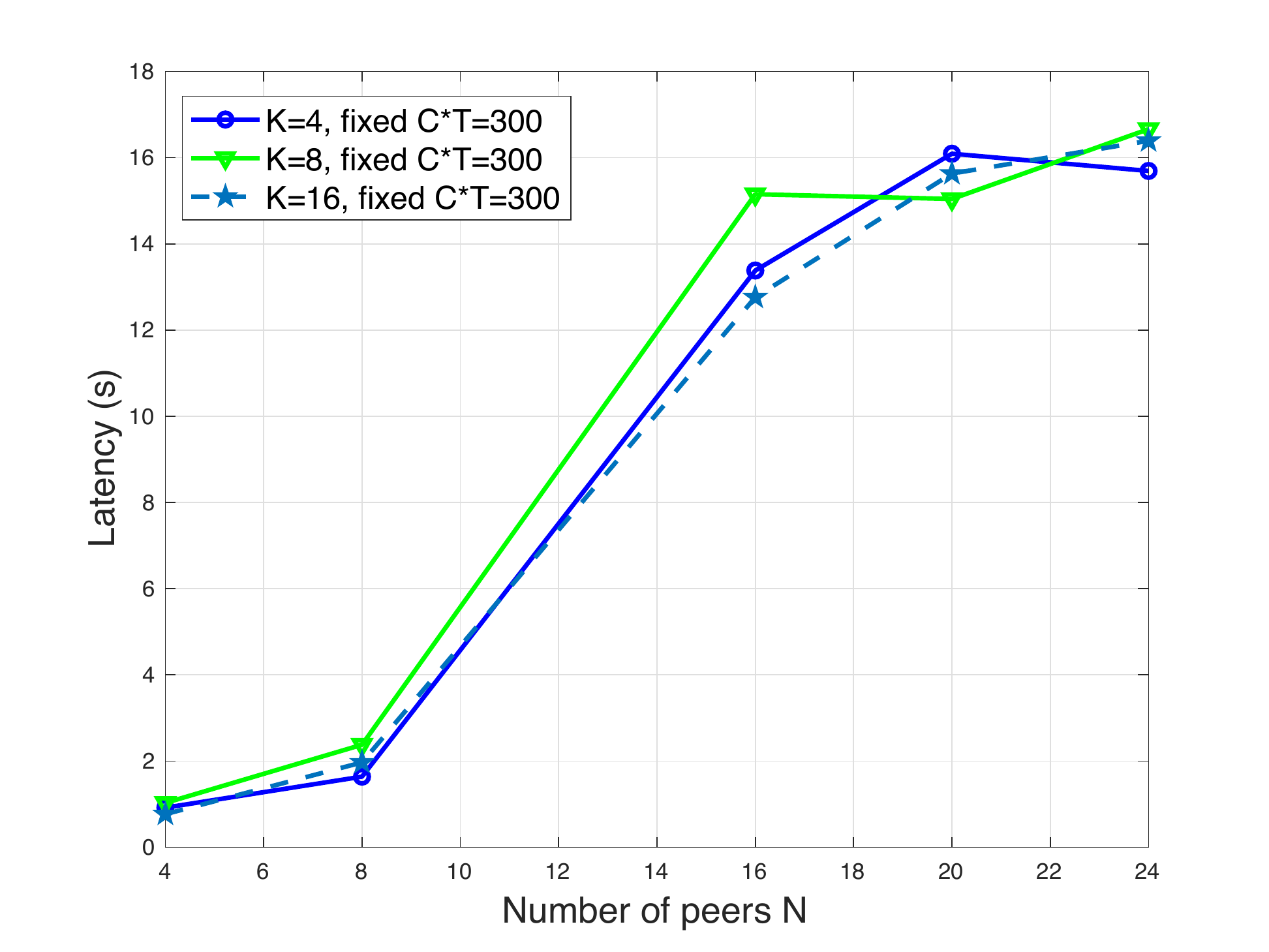}
    \caption{Latency for fixed request rate $C*T=300$ tps}
    \label{scaling_peer_latency}
    \end{subfigure}    
    \begin{subfigure}{0.49\textwidth}
    \includegraphics[width=\textwidth]{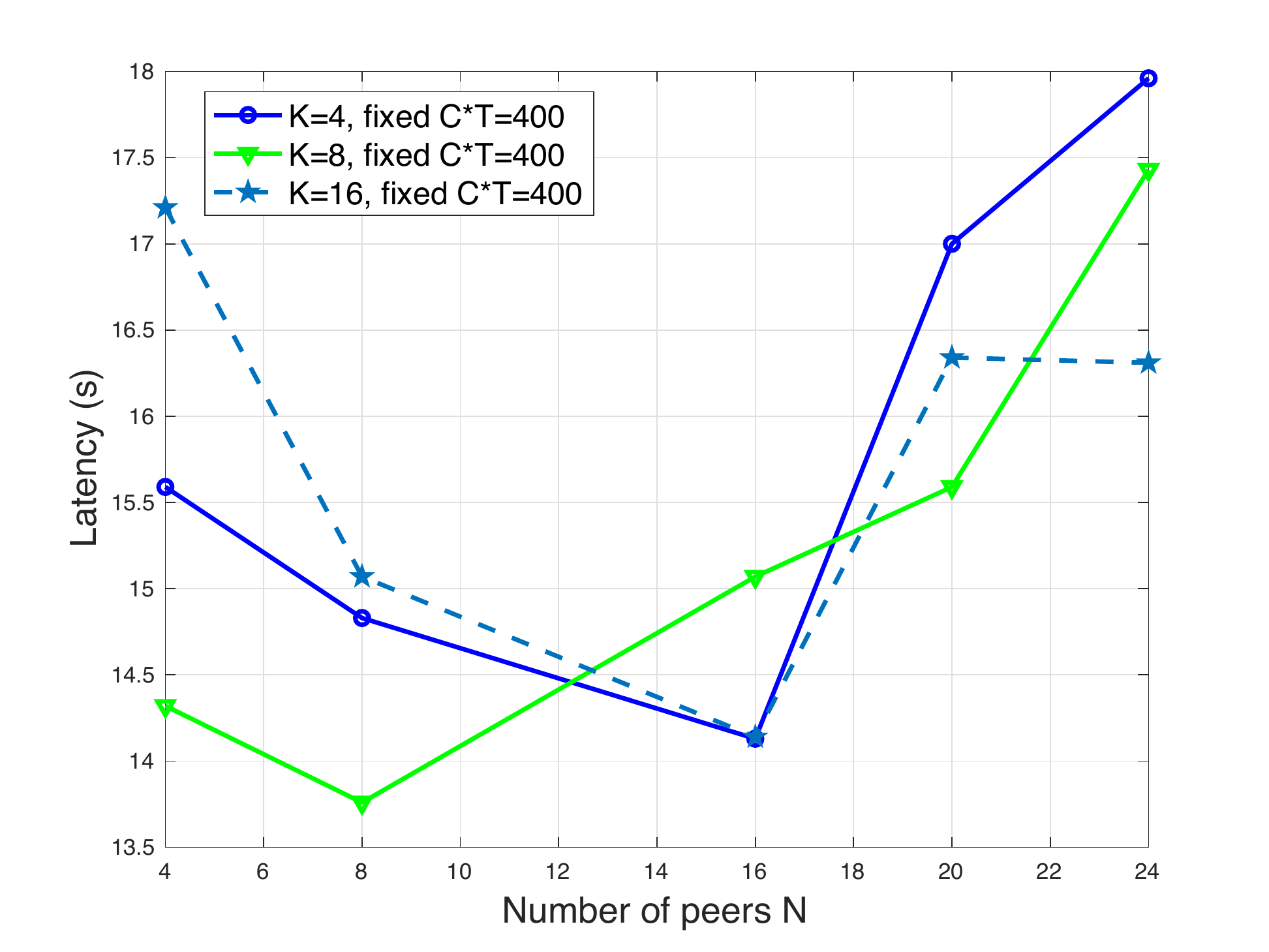}
    \caption{Latency for fixed request rate $C*T=400$ tps}
    \label{scaling_peer_highlatency}
    \end{subfigure}
    \caption{Latency while scaling the number of peers}
    \label{scaling_peer_latency}
\end{figure}

Secondly, the orderers broadcast blocks to all the endorsing peers for
validation. Hence, more peers correspond to more communication overhead for the
ordering service. In the experiment with a fixed request rate of 400 tps and
$N=24$ peers, we observe that Caliper's clients start to return timeout errors
while sending endorser transactions to orderers. Further examining the orderers'
logs shows that the rate at which orderers receive endorser transactions from
clients dominates the rate of returning confirmed transactions in the form of
blocks. In particular, whenever an orderer broadcasts a transaction to the Kafka
cluster, it logs a message \textit{"[channel: mychannel] Enqueueing
envelope\ldots "}. After the transaction is ordered by the Kafka and returned to
the orderer for batching blocks, the orderer logs a message \textit{"[channel:
mychannel] Envelope enqueued successfully"}, followed by the immediate
acknowledgement from the client that have just sent such transaction,
\textit{"[channel:
mychannel] Broadcast has successfully enqueued message of type
ENDORSER\_TRANSACTION from\ldots"}. We count the number of such messages and
compute the ratio $r = \frac{\text{number of "Enqueueing envelope..." messages
}}{\text{number of "Envelope enqueued successfully" messages}}$. On average,
this ratio is larger than 1.7, suggesting that the Kafka-based ordering service
can only order and return to the commit phase $1/1.7\approx 58.8\%$ of the
amount of endorser transactions it receives.
As increasing the number of peers only lowers the endorser transaction rate sent
to orderers, as explained in the previous paragraph, and there is no interaction
between peers and Kafka cluster, the bottleneck of orderers totally lies in the
communication overhead for broadcasting to a large number of peers.

Letting aside these issues, the scalability for Hyperledger Fabric v1.1 improved
against v0.6. Blockbench \cite{blockbench} shows that v0.6
fails with more than 8 peers. However, the current performance of Hyperledger
Fabric v1.1 is still limited in terms of real-world applications. For example,
other two popular blockchain systems, Ethereum and Parity, scale up to more than
32 nodes while maintaining the steady throughput around 100 and 50 tps
respectively \cite{blockbench}.

\subsection{Communication Overhead within Kafka Cluster}

In this experiment, we fix $N=K=C=16$ and investigate the effect of
communication within the Kafka cluster on Fabric's throughput, by comparing two
extreme settings. In the Kafka-based Fabric ordering service, a transaction must
be written to $default.replication.factor$ number of Kafka brokers, and
committed after $min.insync.replicas$ number of successful writes. In our first
setting, we set these two values to their maximum, $default.replication.factor =
15$ and $min.insync.replicas = 14$, to cause a high communication cost inside
the Kafka cluster. In the second setting, we set these two values to the minimum
possible, $default.replication.factor = 1$ and $min.insync.replicas = 1$, while
violating fault-tolerance. Under the least communication setting, a transaction
is committed immediately after being written to one Kafka broker.
After that, the Kafka cluster still writes the transaction to the remaining
Kafka brokers, but the Fabric's transaction flow does not halt for this
operation, thereby incurring minimal communication overhead from Kafka service.
Therefore, the gap between these two cases relatively represents the overhead of
the Kafka ordering service under full fault-tolerance, represented by
$default.replication.factor = K - 1$.

\begin{figure}[t]
  \centering
    \includegraphics[width=0.5\textwidth]{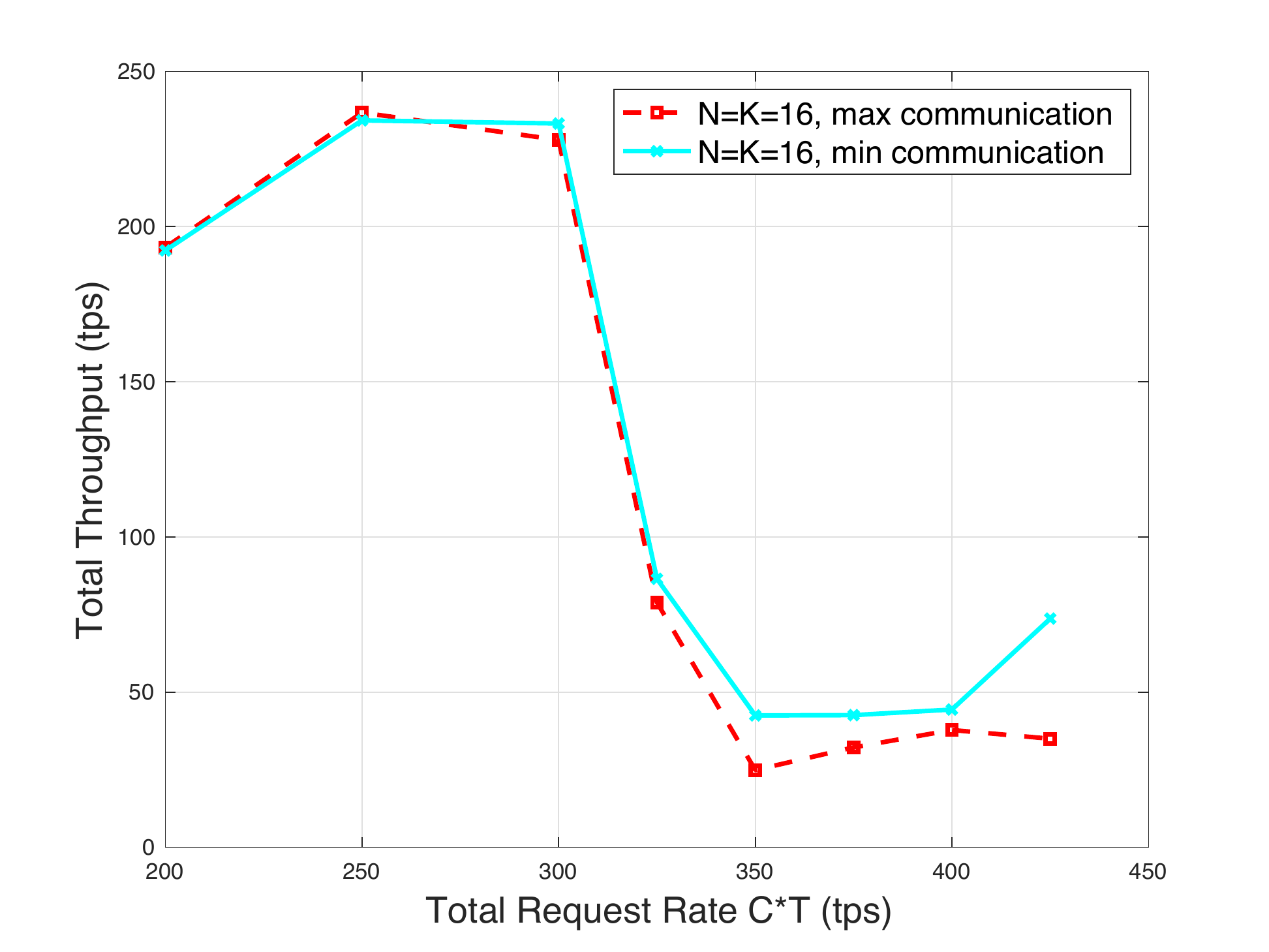}
  \caption{Kafka communication test}
  \label{kafka_heavy}
 \end{figure}

The results, depicted in the Figure \ref{kafka_heavy}, show that the throughput
of the most intensive communication setting is always dominated by the
throughput of the least communication setting. However, as the difference is
negligible, we conclude that intra-cluster Kafka communication does not affect
the overall throughput of Fabric.

\subsection{Scaling Kafka brokers}

In this set of experiments, we fix the number of peers $N$, and investigate the
effect of increasing the number of Kafka brokers. We fix the total request rate
to $C*T=300$ and $C*T=400$. For each value of $N \in \{4, 8, 16\}$,
followed by the same number of clients, we vary the number of Kafka brokers. The
average throughput and latency of each setting is shown in Figure
\ref{scaling_kafka_throughput} and Figure \ref{scaling_kafka_latency},
respectively.

Generally, scaling the Kafka brokers does not impact the throughput pattern or
scalability of the system.

\begin{figure}[t]
    \centering
    \begin{subfigure}{0.49\textwidth}
    \includegraphics[width=\textwidth]{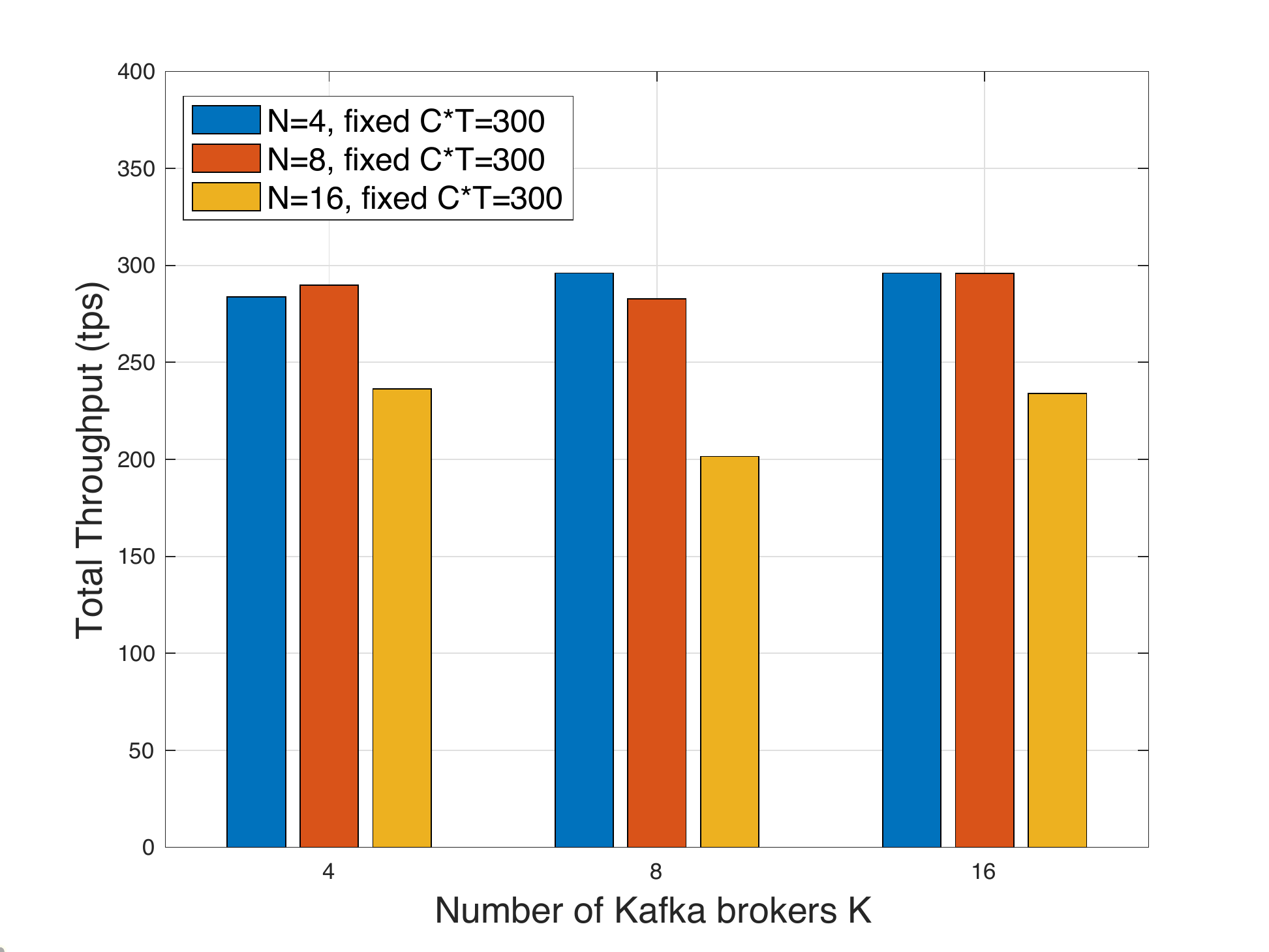}
    \caption{Throughput for request rate $C*T=300$ tps}
    \label{scaling_kafka_throughput}
    \end{subfigure}
    \begin{subfigure}{0.49\textwidth}
    \includegraphics[width=\textwidth]{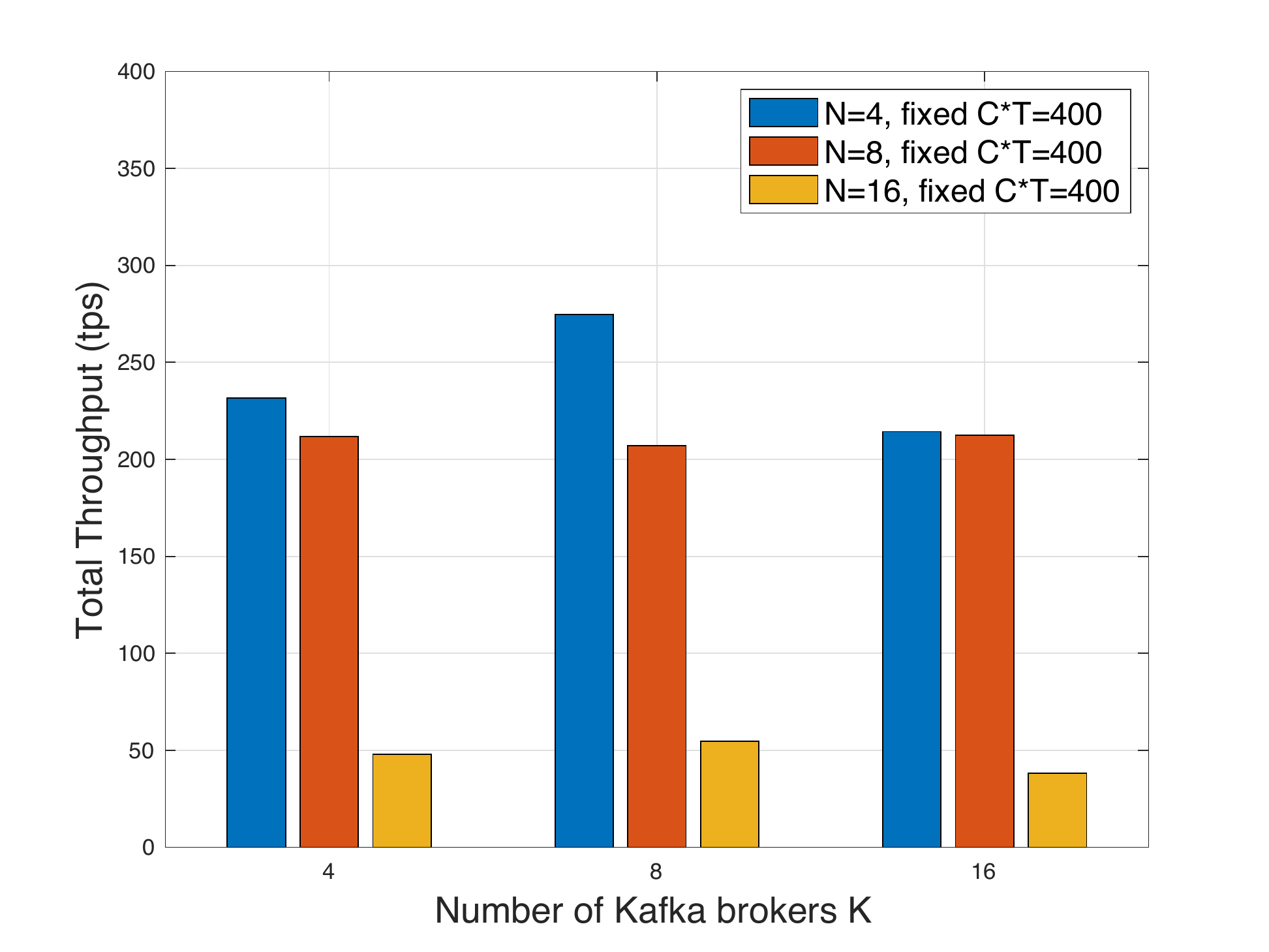}
    \caption{Throughput for request rate $C*T=400$ tps}
    \label{scaling_kafka_highthroughput}
    \end{subfigure}
    \caption{Throughput Performance}
    \label{scaling_kafka_throughput}
\end{figure}

\begin{figure}[t]
    \centering
    \begin{subfigure}{0.49\textwidth}
    \includegraphics[width=\textwidth]{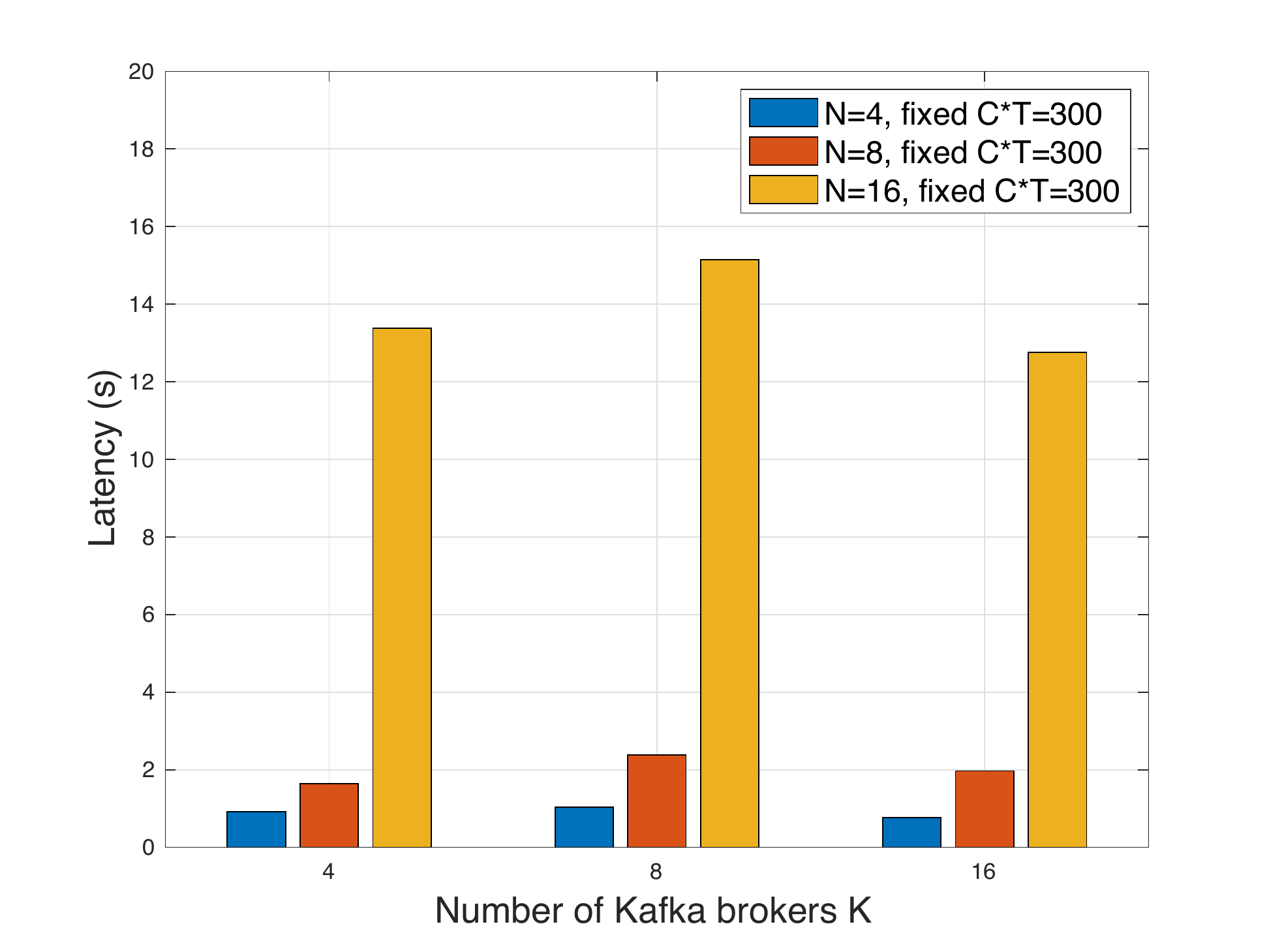}
    \caption{Latency for request rate $C*T=300$ tps}
    \label{scaling_kafka_latency}
    \end{subfigure}
    \begin{subfigure}{0.49\textwidth}
    \includegraphics[width=\textwidth]{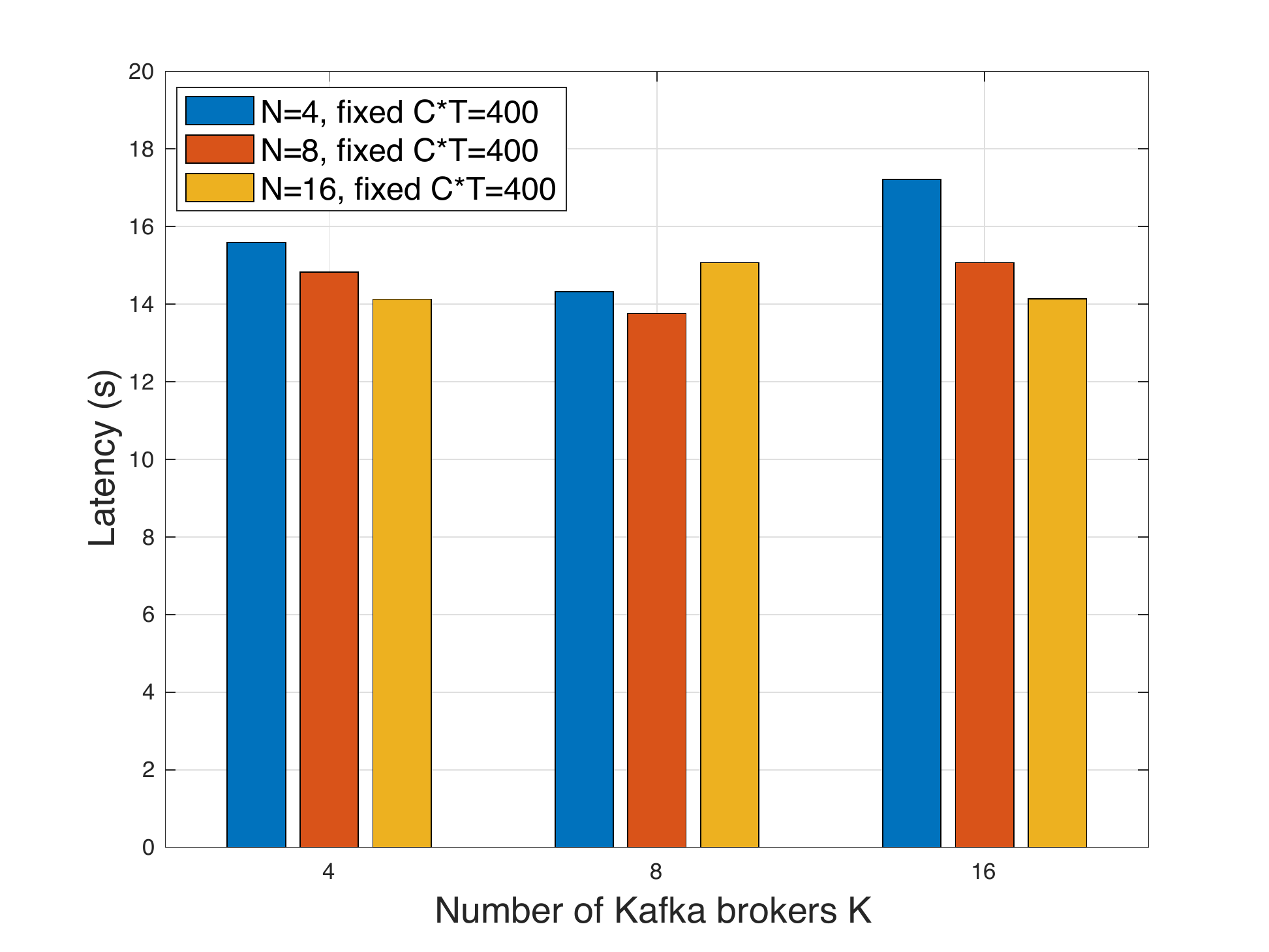}
    \caption{Latency for request rate $C*T=400$ tps}
    \label{scaling_kafka_highlatency}
    \end{subfigure}
    \caption{Latency Performance}
    \label{scaling_kafka_latency}
\end{figure}

\section{Conclusion} 
\label{sec:conclusion}

In this work, we presented our benchmarking tool Caliper++, which is
specifically designed to examine the scalability of Hyperledger Fabric v1.1, a
permissioned blockchain platform. Using Caliper++, we have conducted a
comprehensive study on the scalability performance of Fabric. By identifying
major bottlenecks of  system,  we hope that the survey and benchmarking tool
would serve to guide the design and implementation of Hyperledger Fabric in the
future.

\bibliographystyle{abbrv}
\bibliography{references}

\end{document}